%% file: jimis.tex
\documentclass{jimis}
\usepackage{array}
\usepackage{graphicx}
\usepackage[english]{babel}
\usepackage[utf8]{inputenc}
\usepackage[left,modulo,pagewise]{lineno}
\usepackage{amssymb}
\usepackage{amsmath}
\makeatletter
\@fleqnfalse
\@mathmargin\@centering
\makeatother
\usepackage{graphicx}
\usepackage{adjustbox}

\usepackage[algoruled,noend,linesnumbered]{algorithm2e}
\SetAlFnt{\footnotesize}

\usepackage{textcomp}
\usepackage{enumitem}
\usepackage{xtab}
\usepackage{tabularx}
\usepackage{array}
\usepackage{url}

\usepackage{breakurl}
\title{Brazilian Congress structural balance analysis}
\author[*1]{Mario Levorato}
\author[1]{Yuri Frota}
\affil[1]{Department of Computer Science, Fluminense Federal University, Brazil} 
\corrauthor{mlevorato@ic.uff.br}

\begin{document}

\maketitle

%\linenumbers

\abstract{In this work, we study the behavior of Brazilian politicians and political parties with the help of clustering algorithms for signed social networks. 
For this purpose, we extract and analyze a collection of signed networks representing voting sessions of the lower house of Brazilian National Congress. We process all available voting data for the period between 2011 and 2016, by considering voting similarities between members of the Congress to define weighted signed links. The solutions obtained by solving Correlation Clustering (CC) problems are the basis for investigating deputies voting networks as well as questions about loyalty, leadership, coalitions, political crisis, and social phenomena such as mediation and polarization.}


\section{Introduction}
\input{introduction.tex}

\section{Related works} \label{sec:problem}
\input{problems.tex}

\section{Network extraction}
\input{extraction.tex}

\section{Structural balance analysis}
\input{results.tex}

\section{Concluding remarks}
\input{conclusion.tex}

\bibliographystyle{jimis}
\bibliography{SG}

%\appendix\footnotesize

%\section{Acknowledgements}
%Dans Bibtex, comment écrire une citation d'un article de JIMIS sans rien oublier et dans le bon format? Voir la structure dans les commentaires de \textit{jimis.tex} (annex 2).

% @incollection{jimis,
%   author = {Jimis, Jimmy},
%   title = {Cross-interdisciplinary},
%   booktitle = {Journal of Interdisciplinary Methodologies and Issues in Science},
%   publisher = {Episciences},
%   year = {2015},
%   editor = {James Jamy and Jane Jany},
%   volume = {0},
%   pages = {1-25},
%   edition = {A first Issue about Interdisciplinary Methodologies and Issues in Science},
%   doi = {XXXX-000},
%   note = {\url{http://jimis.episciences.org}}
%  }
 
\end{document}

%% file: introduction.tex
Structural balance theory is based on the notion of cognitive consistency between friendship and hostility. For example, an enemy of a friend is probably my enemy as well, while a friend of a friend is probably my friend or can become one~\citep{Heider46}. In simple terms, the interaction of individuals follows the tendency to create stable (albeit not certainly conflict-free) social groups. This can be specially interesting to study similarity and correlation networks, like those originated from common voting patterns, or alliances and disputes among parties or nations~\citep{Traag09,Macon12,doreian2015structural}.

One appropriate criterion to measure the degree of balance in signed social networks is by solving the Correlation Clustering (CC) problem~\citep{bansal02,demaine06}, which consists of partitioning a set of elements into clusters by analyzing the level of similarity between them. It aims to maximize the affinity inside each cluster (i.e. positive relationships) while, at the same time, minimizing the similarities between elements of different clusters (i.e. maximizing negative relationships). 

The CC problem, which has been proved to be NP-hard~\citep{bansal02}, can be applied in several areas, such as efficient document classification~\citep{bansal02}, natural language processing~\citep{Elsner09}, image segmentation~\citep{kim2014image} and, of course, signed social network analysis~\citep{doreian96,Brusco11,Figueiredo12SN,SAC2015}. 
With this objective, the level of balance in a social group can be used by social network researchers to study how (and if) a group evolves to a possible balanced state.

A relaxed version of the CC problem called Symmetric Relaxed Correlation Clustering (SRCC) problem~\citep{Brusco11,Figueiredo12SN} can also be used to evaluate balance in signed social networks. 
This variant, although computationally harder to solve, allows the identification of special types of social relationship, such as polarization, mediation and differential popularity~\citep{doreian09}, originally viewed as violations of structural balance. 

We implemented an algorithm known as $ILS-CC$~\citep{SAC2015}, which can efficiently solve the aforementioned problems, providing useful information for social network analysis. Using the House of Cunha website~\citep{housecunha} and the work of~\cite{israel2015parliament} as inspiration, we provide a novel analysis of Brazilian politics inside the Chamber of Deputies (CD). In Brazil, the Chamber of Deputies (\textit{C\^amara dos Deputados}) is the lower house of the National Congress, comprised of 513 federal deputies (from 25 political parties), elected by a proportional representation of votes to serve a four-year term. 
Based on the CD voting records, we generate several instances of signed social networks, according to certain grouping criteria. 
The clustering results obtained when invoking the $ILS-CC$ procedure over these instances is the starting point of our study.

% Nossa analise pode ser aplicada em qualquer rede social que represente grupos politicos
The analysis presented in this work can be applied to any network originated from voting patterns, where alliances and interest groups have strong influence. 

This paper is organized as follows. Section II presents a literature review regarding Correlation Clustering problems and signed social network analysis. Section III describes the method applied to extract signed networks from the Chamber of Deputies voting data. %Section IV summarizes the algorithm used to partition these signed networks. 
Section IV presents an analysis of structural balance on the Chamber of Deputies voting networks, based on the solutions obtained by using our methodology. Finally, we show our conclusions in Section V.

%Section 2 presents the Correlation Clustering problem, including a mathematical formulation and a literature review. Section 5 describes the SRCC problem, its application and an efficient solution method. 
%Our analysis relies on the results obtained when solving two clustering problems over these instances, taking negative relationships into account.

%% file: problems.tex
\cite{Heider46} was the first to state Structural Balance (SB) theory in order to define sentiment relations among people belonging to the same social group (such as like/dislike, love/hate and cooperation/competitivity). Signed graphs were later applied by \cite{Cartwright56}, formalizing SB theory which affirmed that a stabilized social group could be divided into two mutually
hostile subgroups (or clusters), each having internal solidarity. \cite{Davis67} then proposed the more general notion of "weak balance" or clusterable signed graph, when a balanced social group can be divided into two or more mutually antagonistic subgroups, each having internal solidarity.

When solving a clustering problem, one wants to find the most balanced partition\footnote{A partition is here defined as the division of the set of vertices $V$ into non-overlapping and non-empty subsets.} of a signed graph. Using structural balance as a measure, the clustering problem is equivalent to solving the optimization problem called Correlation Clustering (CC). To our knowledge, this problem was first addressed by \cite{doreian96} (although not under this name), who provided a heuristic solution method for analyzing structural balance on real-world social networks. 
Their method was implemented in software Pajek~\citep{PajekWiki}. 
Having a document clustering problem in mind, 
\cite{bansal02} formalized the unweighted version of the CC problem and also discussed its NP-completeness proof. Later, \cite{demaine06} addressed the weighted version of the problem.
Integer linear programming (ILP) can be used to solve the CC problem optimally~\citep{Figueiredo12SN}, but only if the number of elements is small. 
Since it consists of a NP-hard minimization problem, the only available solutions for larger instances are either heuristic or approximate. 
%Predominately investigated from the viewpoint of constant factor approximation algorithms \citep{bansal02,swamy2004correlation,Charikar05, demaine06, giotis2006correlation,ailon2008aggregating}, 
The solution of the CC problem and of some of its variants has already
been applied in several areas, % Melhorar essa parte
such as portfolio analysis in risk management~\citep{Harary03, huffner09}, biological systems~\citep{DasGupta07,huffner09}, grouping of genes~\citep{bhattacharya2008divisive}, efficient document classification~\citep{bansal02}, image segmentation~\citep{kim2014image} and community structure~\citep{Traag09}.

%Additionally, some authors have focused on the detection of overlapping communities in signed graphs~\cite{zhang2007identification, bonchi2011overlapping}. In particular, \cite{bonchi2011overlapping} define an optimization problem that extends the framework of Correlation Clustering to allow overlaps.

%An evaluation of different heuristic strategies (greedy and local search methods) for the CC problem is performed in~\cite{Elsner09}, using two practical tasks: document clustering\footnote{The original data set from UCI Machine Learning Repository \citep{Bache+Lichman:2013} consists of $20000$ messages taken from $20$ newsgroups. However, since the authors' bounding technique did not scale to the full dataset, they restricted their testbed to a subsample of $100$ messages from each newsgroup, for a total of $2000$ messages.} and natural language processing (instances of $n=1000$), to which ILP does not scale. 
%In this situation, the authors' recommendation for CC Problem solution is the $VOTE/BOEM$ greedy algorithm, which rapidly obtains good objective values (with tight bounds). 
%A greedy neighborhood-based heuristic for the problem has also been proposed by \cite{wang2013restoring}. Instead of stepping through the vertices according to a vertex permutation, their algorithm, known as Restoring, partitions vertices according to an edge permutation.

In~\cite{Yang07}, the CC problem is known as {\it community mining} and an agent-based heuristic called FEC is proposed to obtain its solution. 
%In order to assess the performance of FEC, the authors present a method for generating random signed networks with controlled community structures, based on a set of community structure-conscious parameters. 
Genetic algorithms have also been applied to document clustering, using the CC problem as objective function~\citep{Zhang08}. 
Lately, ~\cite{DrummondSOMOCO2013} presented a Greedy Randomized Adaptive Search Procedure (GRASP) \citep{feo1995grasp} implementation that provides an efficient solution to the CC problem in networks of up to $8000$ vertices. 
Then, based on this method, \cite{SAC2015} introduced sequential and parallel ILS (Iterated Local Search) \citep{lourencco2003iterated} procedures for the CC problem (known as $ILS-CC$), which outperformed other solution methods from the literature on three huge real-world signed social networks. Similarly to \cite{israel2015parliament}, in this work, we will use the $ILS-CC$ algorithm to evaluate the imbalance of voting networks.

% FALAR DO SRCC PROBLEM AQUI
Apart from the CC problem, alternative measures to structural balance and the associated clustering problems have also been discussed in the literature. In~\cite{doreian09}, the definition of a $k$-balanced signed graph was informally extended in order to include relevant processes (polarization, mediation, differential popularity and subgroup internal hostility) that were originally viewed as violations of structural balance. 
For example, the existence of a group of individuals who share only positive relationships with everyone in the network counts as imbalance in the CC Problem. Nonetheless, the individuals in this group could be identified as mediators (i.e. their relations probably won't change over time) and, as pointed in \cite{doreian09}, their relations should not be considered as a contribution to the imbalance of the network.

Using this new definition, structural balance was generalized to a version labeled as {\it relaxed structural balance}~\citep{doreian09}. Similarly to the CC problem, measuring the relaxed structural balance can be accomplished through the solution to the Relaxed Correlation Clustering (RCC) problem. It is originally defined on asymmetric relations between clusters~\citep{Figueiredo12SN}; however, a redefinition of relaxed imbalance of a partition $P$ that takes into account only symmetric relationships is also available. 
This gives rise to a new graph clustering problem, the Symmetric Relaxed Correlation Clustering (SRCC) Problem~\citep{Figueiredo12SN}, which will be used in this work. The SRCC problem allows us to analyze mediation processes (positive and negative). That is not the case of the RCC problem, where mediation and differential popularity cannot be pointed out.

It is worth noting that the SRCC problem is closely related with the CC problem but it is not a particular case nor is a generalization. 
Actually, each feasible solution (a graph partition) of the SRCC problem is also feasible in the CC problem but
the problems have different cost functions, i.e., there are different ways of evaluating the imbalance of a partition. 
The SRCC problem is intuitively as difficult as the CC problem and is indeed a NP-hard problem \citep{Figueiredo12SN}. 

Two solution methods were initially presented in the literature for RCC problems: a greedy heuristic approach~\citep{doreian09} and a branch-and-bound procedure~\citep{Brusco11}. Computational experiments with both procedures were reported over literature instances with up to 29 vertices and for random instances with up to 40 vertices~\citep{doreian09,Brusco11}. We extended the ILS procedure to solve the SRCC problem, by applying additional data structures and a new objective function to evaluate the partition~\citep{figueiredo2016ejco}. As far as we know, the $ILS-CC$ algorithm is the only metaheuristic approach that has been used to solve RCC problems.

% ***** REVISAO ESPECIFICA PARA TRABALHOS DE ANALISE DE REDES DE ALIANCAS E DISPUTAS

Previous works have employed signed graph clustering methods to analyze networks of international alliances and disputes~\citep{Traag09,Macon12,doreian2015structural}. 
In~\citet{SAC2015}, by using the $ILS-CC$ algorithm, we presented a historical and geopolitical analysis of the results obtained from the voting on resolutions in the United Nations General Assembly (UNGA). \citet{israel2015parliament} have then applied a parallel version of the $ILS-CC$ algorithm to analyze a collection of signed networks representing voting sessions of the European Parliament. The obtained results were compared to a selection of community detection algorithms designed to process only positive links. 

Several authors studied the voting behavior of politicians. 
As far the European Parliament (EP) is concerned, \cite{hix2009after} compared voting behavior of Members of the European Parliament (MEPs) in different periods, analyzing issues such as party cohesion and coalition formation. On their turn, \cite{hix2002parliamentary} evaluated different questions regarding voting behavior in the EP, including personal policy preferences, national party and European party disciplines.  

In particular, regarding Brazil, \cite{ames1995electoral} developed a model of legislative voting based on the operation of Brazil's political institutions. 
\cite{mainwaring1997party} analyzed party discipline in the Brazilian constitutional congress of 1987–88. 
\cite{figueiredo2000presidential} analyzed how Brazilian presidents have succeeded by relying on the support of disciplined parties in order to get their agendas approved in the Congress. \cite{calvao2015stylized} performed an extensive analysis of data sets available for Brazilian proportional elections of legislators and city councilors throughout the period 1970–2014,
plus a comparative analysis of elections for legislative positions, in different states and years.

%% file: extraction.tex
In this section, we explain the retrieval of raw voting data, and how we extracted signed networks from it.

\subsection{Brazilian Chamber of Deputies}

The Chamber of Deputies (CD) provides web services\footnote{Please visit \url{http://www2.camara.leg.br/transparencia/dados-abertos/dados-abertos-legislativo/webservices}} which supply information about each of its members, including the vote cast by a specific deputy for each proposition evaluated at the CD. A deputy is described by its name, state (one of 27 Brazilian Federative Units) and political party. 

% Tipos de Votos: Sim; Não; Abstenção; Obstrução; Ausência
For a given proposition, a deputy can express his vote in either of
four ways~\citep{glossariocamara}: \textit{Sim} (For: the deputy wants the proposition to be accepted), \textit{Não} (Against: s/he wants the proposition to be rejected), \textit{Abstenção} (Abstain: s/he refuses to take part in the election and does not vote; equivalent to a white vote) and \textit{Obstrução} (Filibuster: a form of obstruction, where debate over a proposition is extended, in order to delay or entirely prevent a vote on the proposal). 

Besides the previous votes, a deputy may not vote at all, which leads to a fifth vote type: Ausência (Absent: the deputy was not present during the voting session). 

% For each proposition, we also have access to the category it belongs to, called Theme (Tema), i.e. the main theme it addresses. All the themes treated during the 2011-2016 period of the CD are listed in Table I, with the number of corresponding propositions.

% TODO INCLUIR TABELA COM A CONTAGEM DE PROPOSICOES POR TEMA

The Chamber of Deputies' web services provide raw voting data, which describe the behavior of deputies apart from the others. Nonetheless, since a network is naturally relational (relationships between individuals are the product of their opinion about topics of interest), voting data has to be processed to generate the networks we wish to analyze.

\subsection{Extraction algorithm} \label{sec:extraction}

The extraction method here explained is based on the work of~\cite{israel2015parliament}. However, this procedure is being applied to Brazilian voting networks for the first time, which demanded an extension to the original algorithm, for filibuster treatment. 
It starts with a comparison between all pairs of deputies, analyzing the similarity of their voting choices. The obtained measures make up what is known as the agreement matrix $M$. Each element $m_{uv}$ of this matrix indicates the average agreement between two deputies $u$ and $v$, in other words, their level of accordance taking into consideration all propositions voted during a given time period. 

While filtering the results is a relatively simple task, processing agreement scores may seriously alter the resulting network, depending on the methodology applied. Given a certain pair of deputies $u$ and $v$ and a proposition $p_i$, the proposition-wise agreement score $m_{uv}$ ($p_i$) is determined by comparing the votes of both deputies. It ranges from -1 if they fully disagree (one voted FOR and the other AGAINST), to +1 if they entirely agree (they share the same vote: FOR or AGAINST). 

% The filtering step is straightforward, however the agreement processing constitutes a major methodological point: depending on how it is conducted, it can strongly affect the resulting network. For a pair of MEPs $u$ and $v$ and a given document $d_i$, we define the document-wise agreement score $m_{uv}$ ($d_i$) by comparing the votes of both considered MEPs. It ranges from -1 if the MEPs fully disagree, i.e. one voted FOR and the other AGAINST, to +1 if they fully agree, i.e. they both voted FOR or AGAINST. 

As previously stated, a voting record may contain, besides FOR and AGAINST, other values which should be equally taken into account. The first case refers to absence of one deputy or both of them (it is worth remembering that the analysis is based on pairs of deputies). The general approach is to leave out all propositions $p_i$ that fall into this case \citep{porter2005network, dal2014voting}. Since certain deputies have low attendance rates, this might lead to distorted agreement or disagreement average scores, due to the small number of common voting sessions. To prevent this, we assume a neutral score of zero if at least one deputy is absent when voting a given proposition. 

%However, as we mentioned previously, a vote can take other values than just FOR and AGAINST, and those must also be treated. The common approach when treating this type of voting data  is to ignore all documents $d_i$ for which at least one of the considered MEPs was absent. However, certain MEPs are absent very often, which mean they would share a very small number of documents with others. This approach could therefore artificially produce extremely strong agreement or disagreement scores. To avoid this, we assign a neutral score of 0 when at least one person is absent for a given document.

The abstention process is more complicated to understand. For example, if the political party supports a completely different view from the deputy, such pressure may be enough to lead him/her to take a step towards abstention, despite the fact that s/he is FOR or AGAINST the proposition under analysis. Similarly, abstention may simply represent the deputy's neutral position when a specific topic is proposed (i.e. the deputy does not care whether or not the subject is approved). Literature provides different views to deal with ABSTAIN-FOR, ABSTAIN-AGAINST and ABSTAIN-ABSTAIN situations \citep{Macon12, porter2005network, dal2014voting}. In this work, we make use of two different ways of calculating the scores. The first one (Table~\ref{tab-vote-weights-half-agree}) treats abstention as half an agreement whenever it is paired with FOR, AGAINST or other abstention, yielding a value of +0.5. In the second one (Table~\ref{tab-vote-weights-absence-opinion}), whenever two deputies abstain at the same time, this is viewed as a full agreement (+1 value). As opposed to that, if only one abstains, a zero score is assigned, since there is not sufficient information to assert they are in agreement or disagreement. So to make things more clear, absence was not included in the tables. 

%Handling the abstentions is a bit trickier, because such a behavior can mean different things. A MEP can choose not to vote because he is personally FOR or AGAINST, but undergoes some pressure (from his political group, his constituents, etc.) to vote the other way: in this case, voting ABSTAIN is a way of expressing this conflicting situation. But abstaining could also simply represent neutrality, meaning the MEP is neither FOR nor AGAINST the considered document. There is no consensus in the literature, and different approaches were proposed to account for ABSTAIN-FOR, ABSTAIN-AGAINST and ABSTAIN-ABSTAIN situations \citep{Macon12, porter2005network, dal2014voting}. Here, we present two variants corresponding to different interpretations. In the first, described in Table II, an abstention is considered as half an agreement with FOR or AGAINST, leading to a score of +0.5. In the second, described in Table III, two abstaining MEPs are considered to fully agree (+1). But, when only one of them abstains, we consider there is not enough information to determine whether they agree or disagree, and we therefore use a 0 score. Note absences were left out of the tables for clarity.

% Este ultimo caso eh novo e nao tinha no trabalho original do Israel
%REVISAR ingles
The last case is filibuster (or obstruction), a vote choice specific to the Brazilian Congress, which does not occur in the European Parliament and was, therefore, not studied by \cite{israel2015parliament}. Such practice is used to create difficulties or hindrances in a systematic way to delay or impede the approval of a bill in parliament. It is normally used by minority groups which do not have the necessary number of representatives to effectively hold back a decision taken by the majority. Therefore, any vote marked as obstruction is here regarded as AGAINST.

\begin{table}[htbp]
\centering
\begin{tabular}{|l|c|c|c|}
\hline
        & \textsc{For}  & \textsc{Abstain} & \textsc{Against} \\ \hline
\textsc{For}     & +1   & +0.5    & -1      \\
\textsc{Abstain} & +0.5 & +0.5    & +0.5    \\
\textsc{Against} & -1   & +0.5    & +1      \\ \hline
\end{tabular}
\caption{Vote weights representing abstention as half an agreement~\cite{israel2015parliament}.}
\label{tab-vote-weights-half-agree}
\end{table}

\begin{table}[htbp]
\centering
\begin{tabular}{|l|c|c|c|}
\hline
        & \textsc{For}  & \textsc{Abstain} & \textsc{Against} \\ \hline
\textsc{For}     & +1   & 0    & -1      \\
\textsc{Abstain} & 0    & +1   & 0    \\
\textsc{Against} & -1   & 0    & +1      \\ \hline
\end{tabular}
\caption{Vote weights representing abstention as absence of opinion~\cite{israel2015parliament}.}
\label{tab-vote-weights-absence-opinion}
\end{table}

The proposition-wise agreement score is fully specified by choosing one of the previous processing strategies. By averaging this score over all considered
propositions, the average agreement can be calculated~\citep{israel2015parliament}. In a formal way, consider two users $u$ and $v$, as well as the propositions resulting from the filtering stage: $p_1$, ..., $p_\ell$, for which both $u$ and $v$ voted. The average agreement $m_{uv}$ between these two deputies is:

\begin{align}
\centering
    %\begin{alignedat}{2}
m_{uv} = \frac{1}{\ell} \sum_{i = 1}^{\ell} m_{uv}(p_i)
    %\end{alignedat}
\end{align}

%Resumindo, foram usadas 2 medidas de similaridade entre os deputados (vide tabela) ambas baseadas na média aritimética do nível de concordância dos votos :
%1) Abstenção vale meia concordância (+0,5);
%2) Abstenção como ausência de opinião (zero);
% Foram gerados 2 grafos por instância de votação, um para cada medida.

% TODO: EXPLICAR ESSE ASPECTO: Ao final da geração dos grafos (inicialmente grafos completos), poderemos torná-los mais esparsos eliminando arestas com peso menor que determinado patamar.

Similarly to the work of~\cite{israel2015parliament}, we generated one signed graph for each year (from 2011 until June 2016), taking into account all the voting sessions in that year. Graph edges with weight smaller than 0.001 were removed from the graph. The set of vertices in each signed graph represents the list of deputies who voted at least one time in the corresponding year.  
%Explicacao geracao de grafos da UNGA
%The set of weighted positive/negative edges is defined as follows. For each pair of vertices (deputies) $i,j$ and for each resolution voted in the session, we totaled the weights associated with all pairs of votes from $i$ and $j$: edge weights can be equal to $1.0$ or $0.5$; a positive edge means an agreement, while a negative edge represents a disagreement. Following an observation from~\citep{Macon12}, we treat differently the disagreement (agreement) in a yes-no (yes-yes or no-no) pair of votes on a same resolution from a yes-abstain or no-abstain (abstain-abstain) pair. We normalize by the total number of votes in a session. 

%% file: results.tex
In this section, based on the clustering results obtained with the ILS-CC algorithm on the graphs extracted according to Section~\ref{sec:extraction}, we investigate some aspects of Brazilian politics in the Chamber of Deputies, including loyalty, leadership, coalition, crisis, as well as social phenomena such as mediation and polarization.

As explained in the previous section, we followed two approaches when generating voting networks for each year in the period between January 2011 and June 2016. We will refer to each network as either v1 or v2, depending on the strategy while dealing with abstentions:
\begin{itemize}
    \item[\textbf{v1}]: abstention is worth half an agreement (+0.5), whenever it is paired with any kind of vote (FOR, AGAINST or other abstention);
    \item[\textbf{v2}]: abstention is viewed as full agreement (+1 value) only if both deputies abstain. Otherwise, if only one abstains, a zero score is assigned.
\end{itemize}

In order to improve the readability of some charts, not all party labels have been displayed. For full information, all charts and tables used in this analysis are available on-line\footnote{Please visit  \url{https://public.tableau.com/profile/mario.levorato} }.

\subsection{A brief introduction to Brazilian politics}

From 1994 to 2002, Brazil was governed by president Fernando Henrique Cardoso, member of the PSDB (Brazilian Social Democracy Party). In 2002, PSDB was defeated in the presidential elections by PT (Brazilian Labor Party) and president Lula da Silva was elected for a four-year term, being reelected in 2006 for one more period of four years. Then, in 2010, president Dilma Rousseff (also a PT member and supported by president Lula da Silva) won the elections, becoming the next president and, like her predecessor, was also reelected in 2014 for an additional four-year term.

Since 2013, Brazil has been facing intense political and economical crisis, aggravated by successive scandals of corruption in the heart of the government~\citep{connors2016, robinsearly2016}. In 2016, an impeachment process was started, on charges related to breaking budget laws, and president Dilma Rousseff was turned away from her post~\citep{watts2016, bbc2016}. However, a more detailed research over international news articles reveals different views about the root causes of the political crisis and the impeachment itself~\citep{alston2016, bevins2016, connors2016, leahy2016, rapoza2016, shahshahani2016, taub2016}.

In order to help understanding the political groups and parties referenced in the analysis, we first provide a list of the three party alliances during the presidential elections held in 2010 (Table~\ref{2010-coalition}) and in 2014 (Table~\ref{2014-coalition}). In our analysis, we will refer to the first party alliance (candidate Dilma Rousseff, in both presidential elections) as the government coalition, while the second party alliance (candidates José Serra in 2010, and Aécio Neves in 2014) will be called opposition.

\begin{table}[htpb]
\centering
\begin{tabular}{|l|l|l|}
\hline
\textbf{Candidate} & \textbf{Coalition parties} & \textbf{\#}                            \\ \hline
Dilma Rousseff                & PCDOB, PDT, PMDB, PR, PRB, PSB, PSC, \textbf{PT}, PTC, PTN & 10 \\ \hline
José Serra              & DEM, PMN, PPS, \textbf{PSDB}, PTB, PTDOB & 6                  \\ \hline
Marina Silva  &  \textbf{PV*}  & 1 \\ \hline
\end{tabular}
\caption{Coalitions in the 2010 presidential elections, ordered by the number of parties (\#) and quantity of votes. Six more candidates (from six remaining parties) ran for presidency in 2010. Like \textbf{PV}, their parties were not in a coalition.}
\label{2010-coalition}
\end{table}

\begin{table}[htpb]
\centering
\begin{tabular}{|l|l|l|}
\hline
\textbf{Candidate} & \textbf{Coalition parties} & \textbf{\#} \\ \hline
Dilma Rousseff       & PCDOB, PDT, PMDB, PP, PR, PRB, PROS, PSD, \textbf{PT} & 9 \\ \hline
Aécio Neves     & DEM, PEN, PMN, \textbf{PSDB}, PTB, PTC, PTDOB, PTN, SD & 9 \\ \hline
Marina Silva  &  PHS, PPL, PPS, PRP, \textbf{PSB}, PSL  & 6 \\ \hline
\end{tabular}
\caption{Coalitions in the 2014 presidential elections, ordered by the number of parties (\#) and quantity of votes. Eight more candidates (from eight remaining parties) ran for presidency in 2014. Their parties were not in a coalition.}
\label{2014-coalition}
\end{table}

Another useful piece of information is the list of parties according to their orientation (Table~\ref{party-orientation}).

% Listar no artigo que partidos são de centro. Explicar que alguns partidos se denominam centro-esquerda ou centro-direita, porém um grande número deles se enquadra no chamado “centrão”.
\begin{table}[ht]
\centering
\resizebox{\textwidth}{!} {
\begin{tabular}{|l|l|l|}
\hline
\textbf{Orientation} & \textbf{Parties} & \textbf{\#} \\ \hline \hline
Left       & PCB, PCDOB, PCO, PSOL, PSTU, PT  & 6  \\ \hline
Center-left     & PDT, PMN, PPL, PPS, PROS, PSB, PSDB, REDE, SD  & 9 \\ 
\hline
Center  & DEM, PEN, PHS, PMB, PMDB, PRP, PSD, PSDC, PSL, PTB, PTC, PTDOB, PTN, PV  & 14 \\
\hline
Center-right    & NOVO, PR, PRB, PSC  & 4 \\
\hline
Right   & PP, PRTB  & 2 \\
\hline \hline
\textbf{Total} & \textbf{-} & \textbf{35} \\ \hline 
\end{tabular} }
\caption{List of Brazilian political parties according to their orientation~\citep{oglobo2016partidos1, oglobo2016partidos2}.}
\label{party-orientation}
\end{table}

Although some parties classify their orientation as center-left or center-right, a great portion of them can be regarded as center parties. 
As of 2016, the block known as "super-center" includes PEN, PHS, PP, PR, PRB, PROS, PSC, PSD, PSL, PTB, PTN and SD. 

As mentioned in the introduction, the Chamber of Deputies (\textit{Câmara dos Deputados}) is the lower house of the National Congress, comprised of 513 federal deputies (from 25 political parties), elected by a proportional representation of votes to serve a four-year term. % TODO INCLUIR UMA TABELA COM O NUMERO DE DEPUTADOS ELEITOS DE CADA PARTIDO, AGRUPADO POR COALISAO
% a table displaying the number of elected deputies from each party/coaliation for each four year term is missing
Table~\ref{elected-deputies-per-party-per-coalition} displays the number of elected deputies from each party/coalition, for the 2010 (2011-2014 term) and 2014 elections (2015-2018 term).

% TODO remover os deputados suplentes dessa lista: manter apenas os eleitos / titulares
% o total de deputados deve dar 513!
\begin{table}[htb]
\centering
  \begin{tabular}{@{}cc@{}}
    % trim={<left> <lower> <right> <upper>}
    \adjustbox{trim={.06\width} {0.29\height} {0.6\width} {.05\height},clip}%
    {\includegraphics[width=1.0\textwidth]{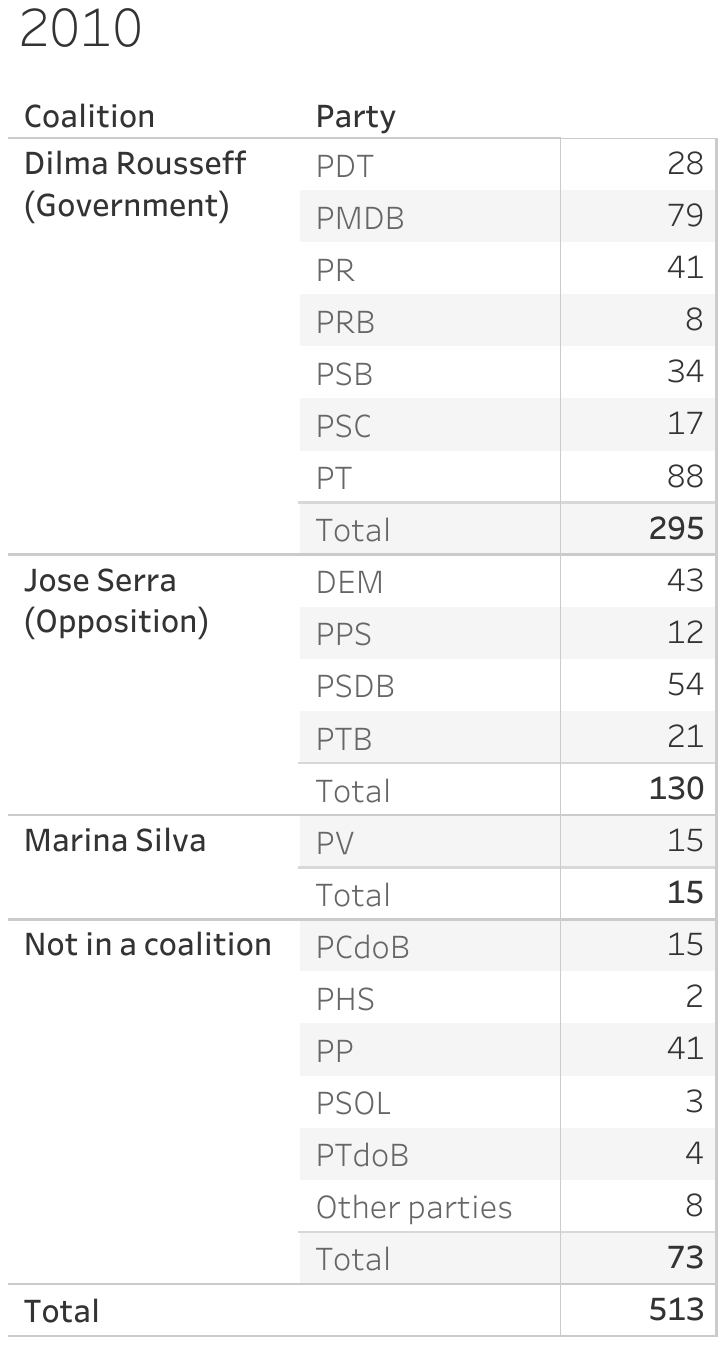}} &
    \adjustbox{trim={.06\width} {0.29\height} {0.6\width} {.05\height},clip}%
    {\includegraphics[width=1.0\textwidth]{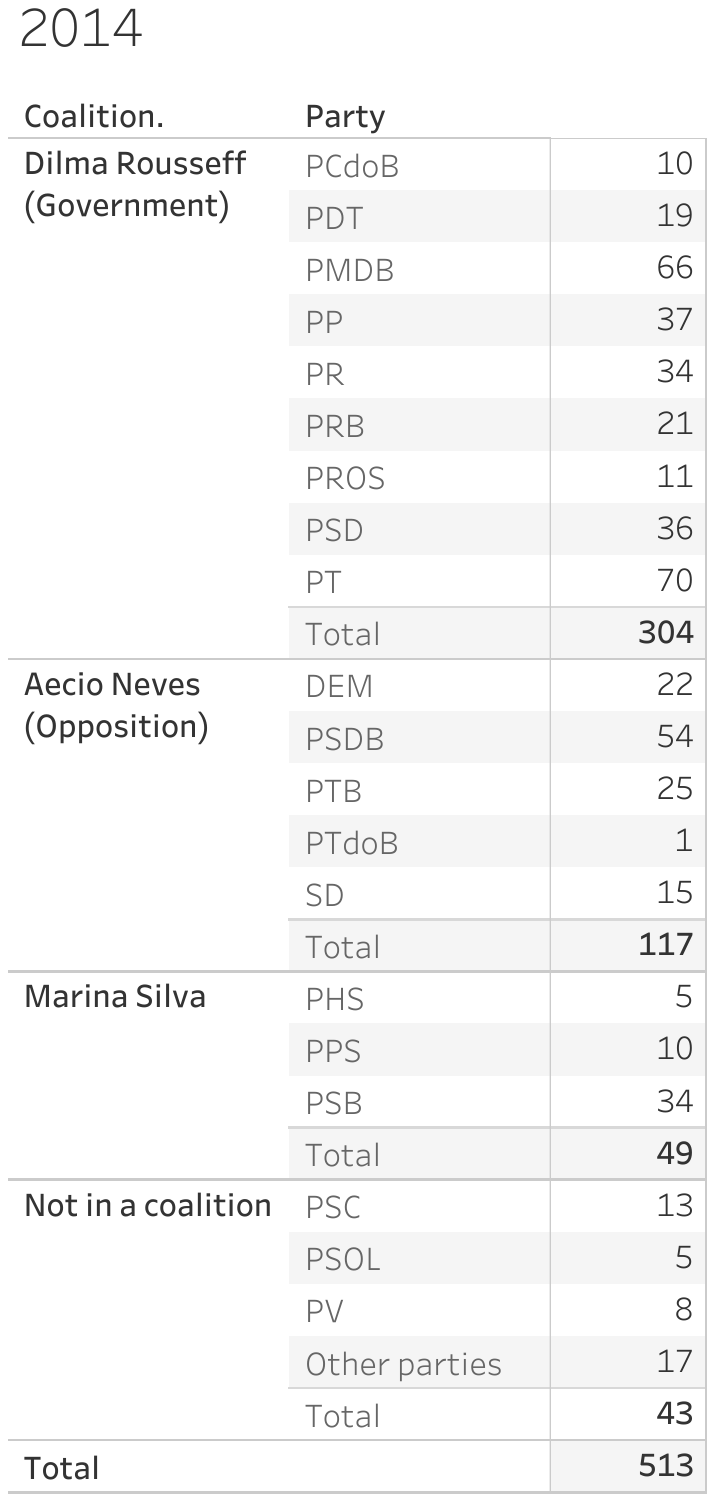}}  \\
  \end{tabular}
  \caption{Number of elected deputies from each party and coalition, for the 2010 elections (2011-2014 term, on the left) and for the 2014 elections (2015-2018 term, on the right).}
  \label{elected-deputies-per-party-per-coalition}
\end{table}

\subsection{Methodology}

We attempt to identify groups of deputies (and their respective parties) in the Chamber of Deputies signed networks, generated based on voting session records publicly made available by the open data initiative of the Brazilian Government~\footnote{The data services of the Brazilian Chamber of Deputies website can be found at \url{http://www2.camara.leg.br/transparencia/dados-abertos}}. 

To do so, we apply the $ILS-CC$~\citep{SAC2015} procedure to solve the two problems introduced in Section II: the Correlation Clustering (CC) problem and the Symmetric Relaxed Correlation Clustering (SRCC) problem. The procedure changes the objective function that evaluates the clustering partition accordingly. 

However, based on the obtained results, we chose to rely our analysis solely on SRCC clustering results~\footnote{We solved the SRCC problem by fixing the number of clusters ($k$) in the solution to $k = 4$, so as to reflect the number of coalition groups: the three main coalitions in each four-year term, listed in Tables~\ref{2010-coalition} and~\ref{2014-coalition}, plus an additional group to represent all the candidates / parties not in a coalition.}. The reason is that all CC solutions presented only one or two clusters as output, which, to our knowledge, did not accurately represent the political groups in the Chamber of Deputies. One possible explanation is that, as stated in Section~\ref{sec:problem}, when compared to the SRCC problem, the CC problem tends to over-evaluate the imbalance of a network, for penalizing relationships associated, for instance, with mediation processes. As we shall see next, parliament mediation groups were indeed detected when solving the SRCC problem. 

Next we present several clustering results that help answering interesting questions concerning political dynamics. Each question and its respective analysis is organized in a subsection. 

% To what extent did parties of the same coalition remain loyal to one another?
\subsection{Evaluation of the loyalty of parties from the same coalition}

We have extracted a table which, for each year, coalition and party (columns \textit{Year}, \textit{Party Alliance} and \textit{Party}, respectively), gives details about the percentage of deputies from each party in each cluster (columns \textit{C1} to \textit{C4}). This way it is possible to spot if the majority of the deputies of a specific party does not belong to the most populous coalition cluster, which constitutes a strong evidence that such party is unfaithful to its coalition. 
By using this data, one can verify that, for example, in 2011 (Table~\ref{tab:CoalitionLoyalty1stGovDetail-2011-v2}), only 41\% of PDT, 38\% of PR and 42\% of PRB deputies were classified inside the largest ruling coalition cluster, formed by 206 deputies. In 2012 (Table~\ref{tab:CoalitionLoyalty1stGovDetail-2012-v2}), only 16\% (3 in 19) of PSC deputies accompanied the biggest government group, comprised of 237 deputies. Finally, in 2014 (on both network versions), just half of PT and PDT deputies followed the government coalition (see column \textit{C1} in Table~\ref{tab:CoalitionLoyalty1stGovDetail-2014-v2}).

\begin{table}[htb]
  \centering
  %\hspace*{-1.0cm}
  % Crop do PDF do Tableau
  % trim={<left> <lower> <right> <upper>}
  \adjustbox{trim={0.07\width} {0.15\height} {0.05\width} {0.12\height},clip}%
  {\includegraphics[scale=0.98]{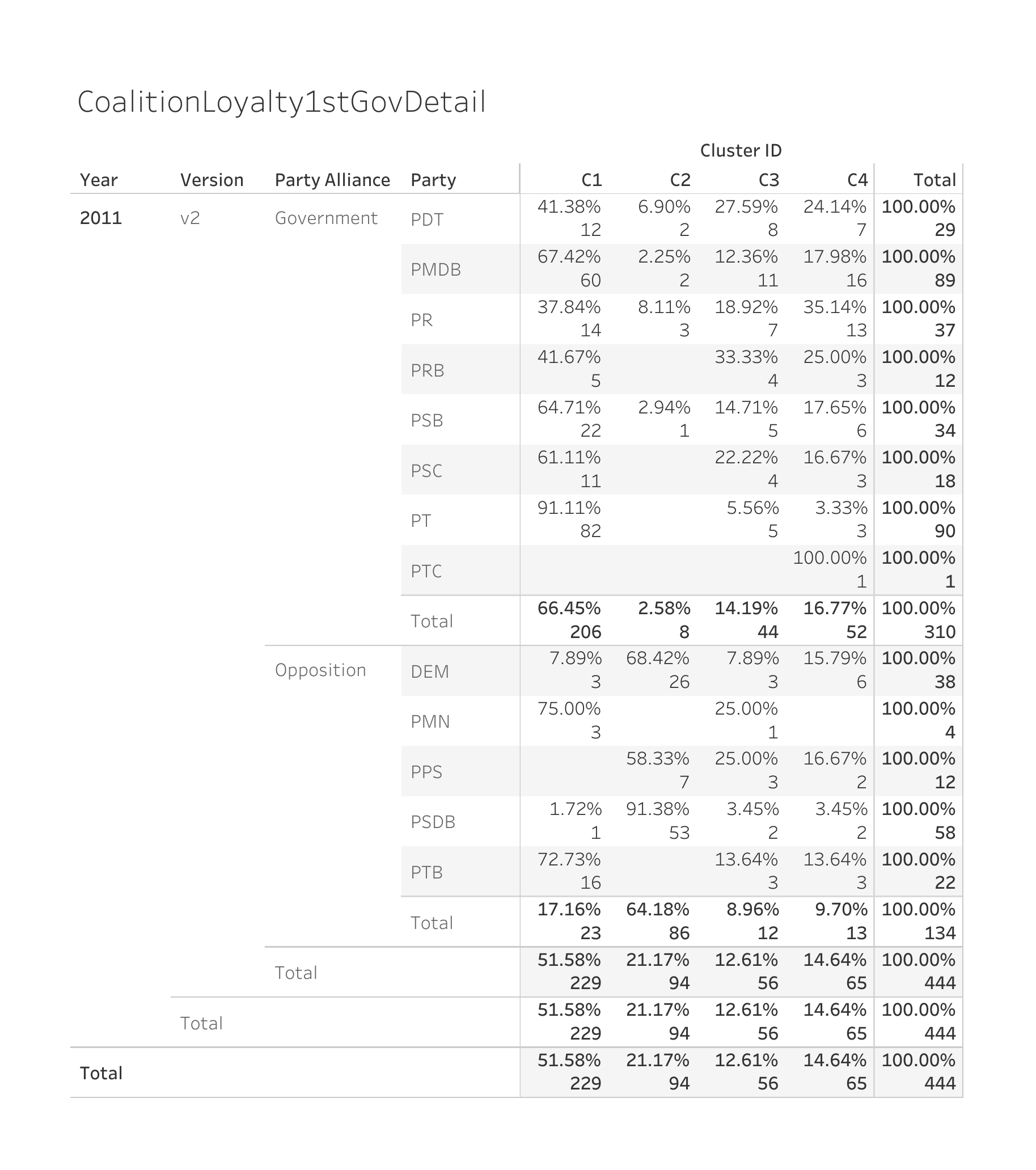}}
  %\vspace*{-1.4cm}
  \caption{Party coalition and clustering details for the year of 2011, when solving version v2 of the voting network, by fixing the number of clusters in the solution to $k = 4$. For each coalition (column Party Alliance) and for each party (column Party), each cell shows the percentage of deputies of that party inside each cluster (columns C1 to C4).}
  \label{tab:CoalitionLoyalty1stGovDetail-2011-v2}
  %\vspace*{-0.5cm}
\end{table}

\begin{table}[htb]
  \centering
  %\hspace*{-1.0cm}
  % Crop do PDF do Tableau
  % trim={<left> <lower> <right> <upper>}
  \adjustbox{trim={0.07\width} {0.15\height} {0.05\width} {0.12\height},clip}%
  {\includegraphics[scale=0.98]{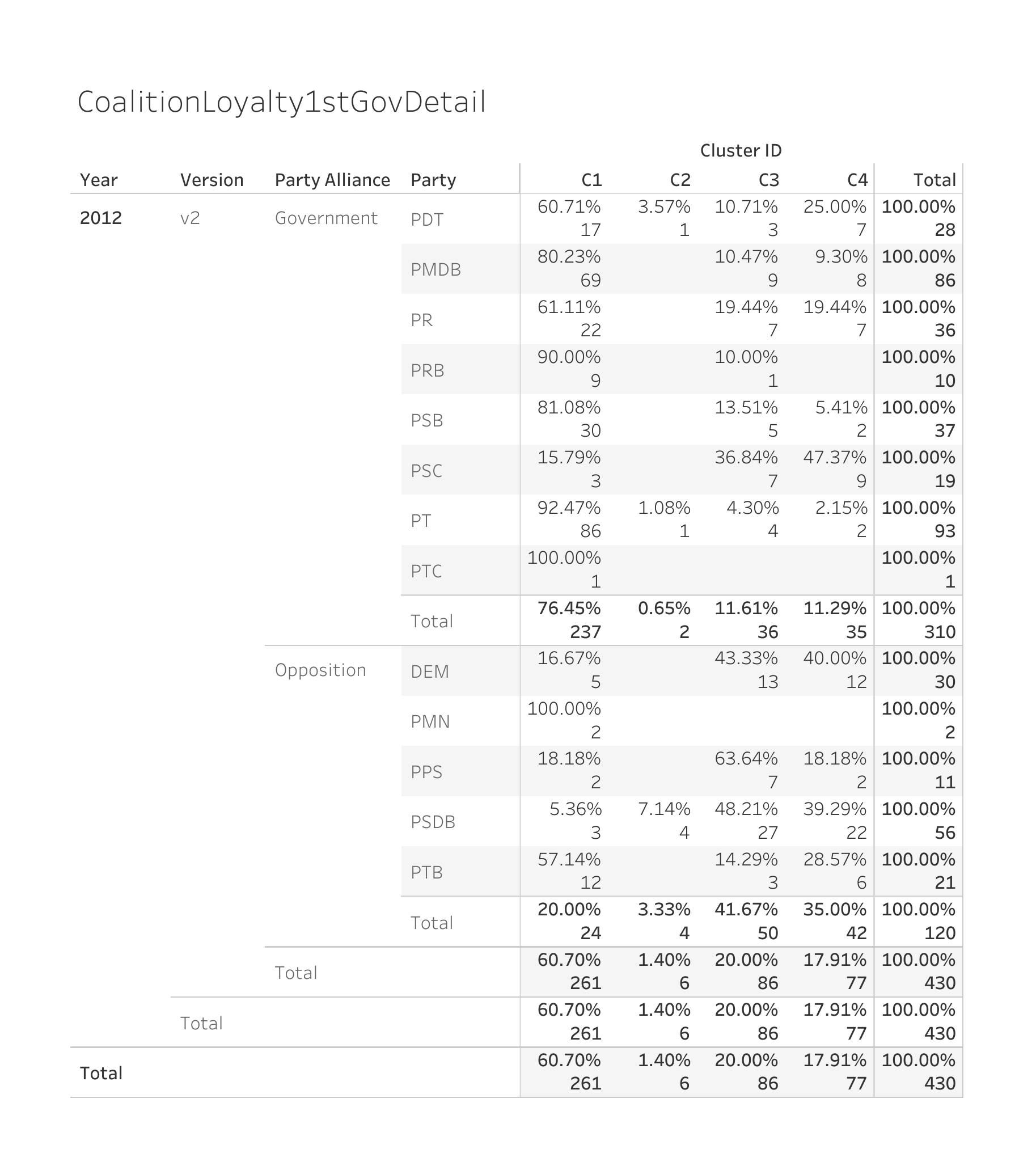}}
  %\vspace*{-1.4cm}
  \caption{Party coalition and clustering details for the year of 2012, when solving version v2 of the voting network, by fixing the number of clusters in the solution to $k = 4$. For each coalition (column Party Alliance) and for each party (column Party), each cell shows the percentage of deputies of that party inside each cluster (columns C1 to C4). }
  \label{tab:CoalitionLoyalty1stGovDetail-2012-v2}
  %\vspace*{-0.5cm}
\end{table}

\begin{table}[htb]
  \centering
  %\hspace*{-1.0cm}
  % Crop do PDF do Tableau
  % trim={<left> <lower> <right> <upper>}
  \adjustbox{trim={0.07\width} {0.15\height} {0.05\width} {0.12\height},clip}%
  {\includegraphics[scale=0.98]{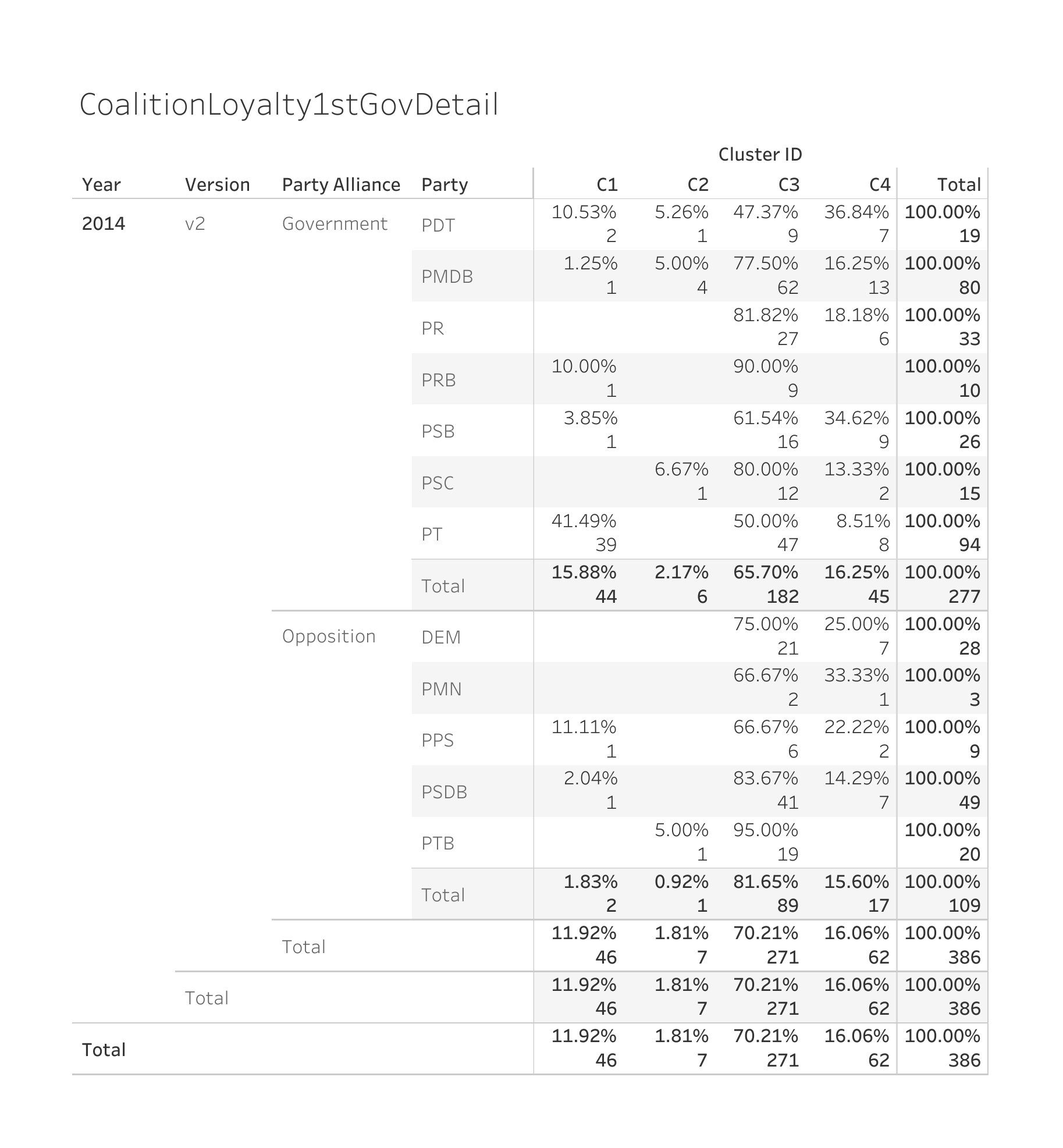}}
  %\vspace*{-1.4cm}
  \caption{Party coalition and clustering details for the year of 2014, when solving version v2 of the voting network, by fixing the number of clusters in the solution to $k = 4$. For each coalition (column Party Alliance) and for each party (column Party), each cell shows the percentage of deputies of that party inside each cluster (columns C1 to C4). }
  \label{tab:CoalitionLoyalty1stGovDetail-2014-v2}
  %\vspace*{-0.5cm}
\end{table}

% Has the government coalition won or lost support?
\subsection{Evolution of the support of the government coalition}

We start by analyzing two tables that provide, for each year and network version (columns \textit{Year} and \textit{Version}, respectively), the number of deputies according to their respective party alliance and the cluster to which they belong (columns \textit{Party Alliance} and columns \textit{C1} to \textit{C4}, respectively). The first table (Table~\ref{tab:CoalitionLoyalty1stGov}) refers to the period from 2011 to 2014 (54th legislature of the Chamber of Deputies), 
while the second one (Table~\ref{tab:CoalitionLoyalty2ndGov}) gives information about the years of 2015 and 2016 (55th legislature, corresponding to president Dilma Rousseff’s second 
presidential term). 

% Modelo para exibicao dos TreeMaps do Tableau
\begin{table}[htb]
  \centering
  %\hspace*{-1.0cm}
  % Crop do PDF do Tableau
  % trim={<left> <lower> <right> <upper>}
  \adjustbox{trim={0.07\width} {0.05\height} {0.05\width} {0.05\height},clip}%
  {\includegraphics[scale=0.9]{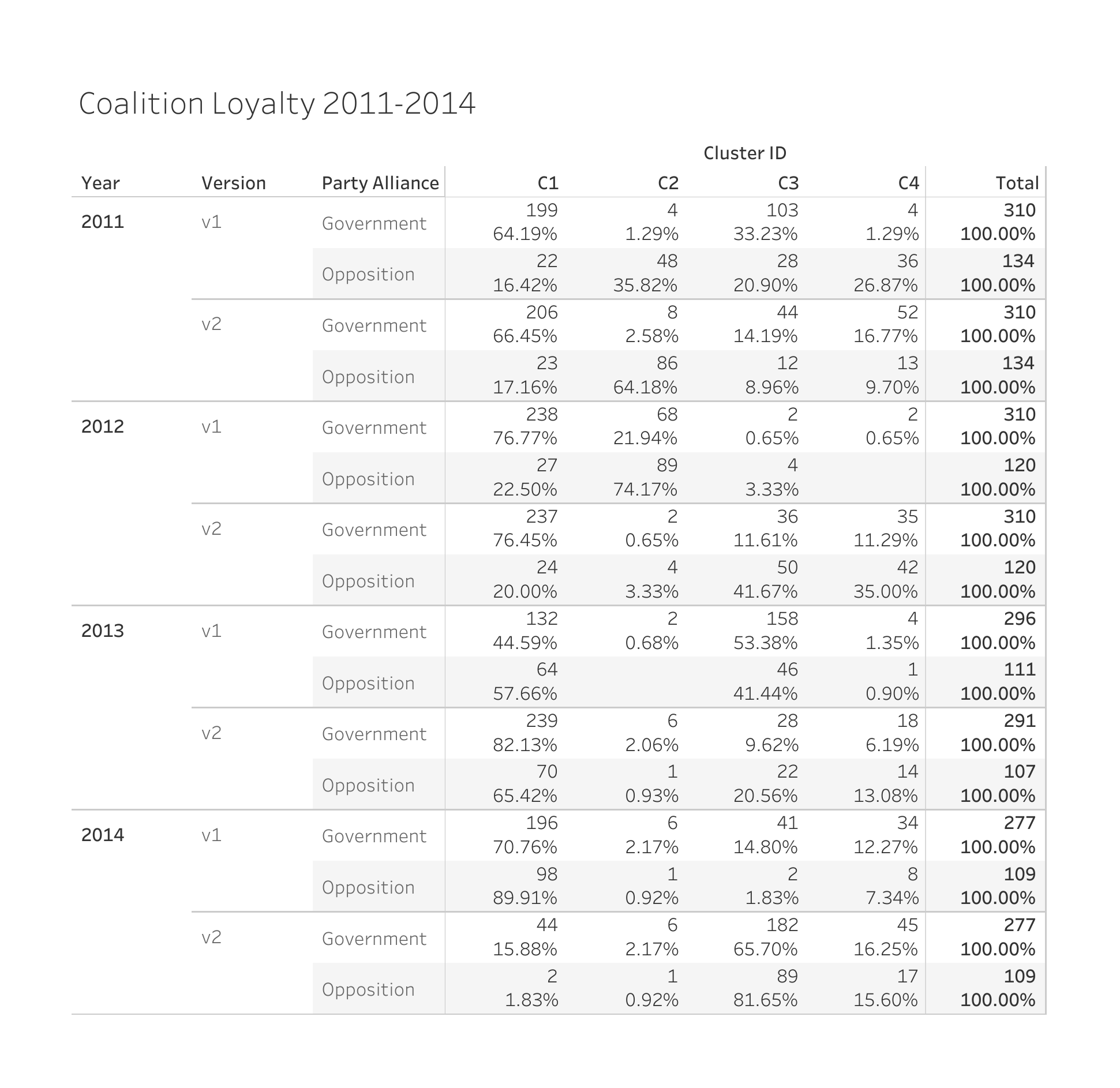}}
  %\vspace*{-1.4cm}
  \caption{Party coalition during the 2010 presidential elections, for the 2011-2014 term. For each year (column \textit{Year}) and network version (column \textit{Version}), the table shows the number of deputies in each party alliance (column \textit{Party Alliance}) found in each cluster (columns \textit{C1} to \textit{C4}). Results obtained when fixing the number of clusters in the solution to $k = 4$.}
  \label{tab:CoalitionLoyalty1stGov}
  %\vspace*{-0.5cm}
\end{table}

\begin{table}[htbp]
  \centering
  %\hspace*{-1.0cm}
  % Crop do PDF do Tableau
  % trim={<left> <lower> <right> <upper>}
  \adjustbox{trim={.07\width} {0.05\height} {0.05\width} {.05\height},clip}%
  {\includegraphics[scale=1.0]{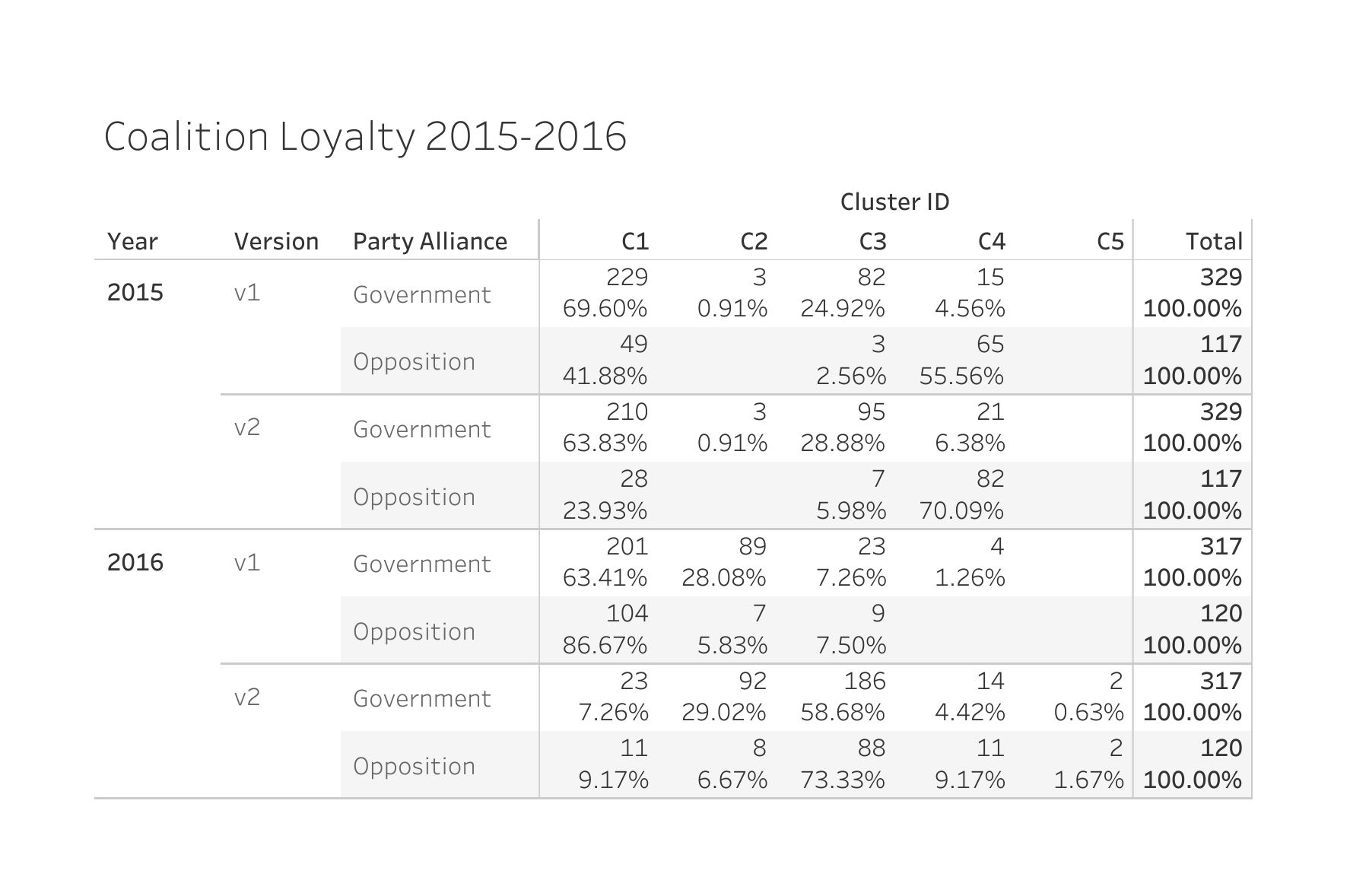}}
  %\vspace*{-1.4cm}
  \caption{Party coalition during the 2014 presidential elections, for the 2015-2018 term. For each year (column \textit{Year}) and network version (column \textit{Version}), the table shows the number of deputies in each party alliance (column \textit{Party Alliance}) found in each cluster (columns \textit{C1} to \textit{C4}). Results obtained when fixing the number of clusters in the solution to $k = 4$.}
  \label{tab:CoalitionLoyalty2ndGov}
  %\vspace*{-0.5cm}
\end{table}

% Modelo para exibicao dos TreeMaps do Tableau
\begin{figure}[htbp]
  \centering
  \hspace*{-1.0cm}
  % Crop do PDF do Tableau
  % trim={<left> <lower> <right> <upper>}
  \adjustbox{trim={.05\width} {.2\height} {0.05\width} {.1\height},clip}%
  {\includegraphics[scale=0.75]{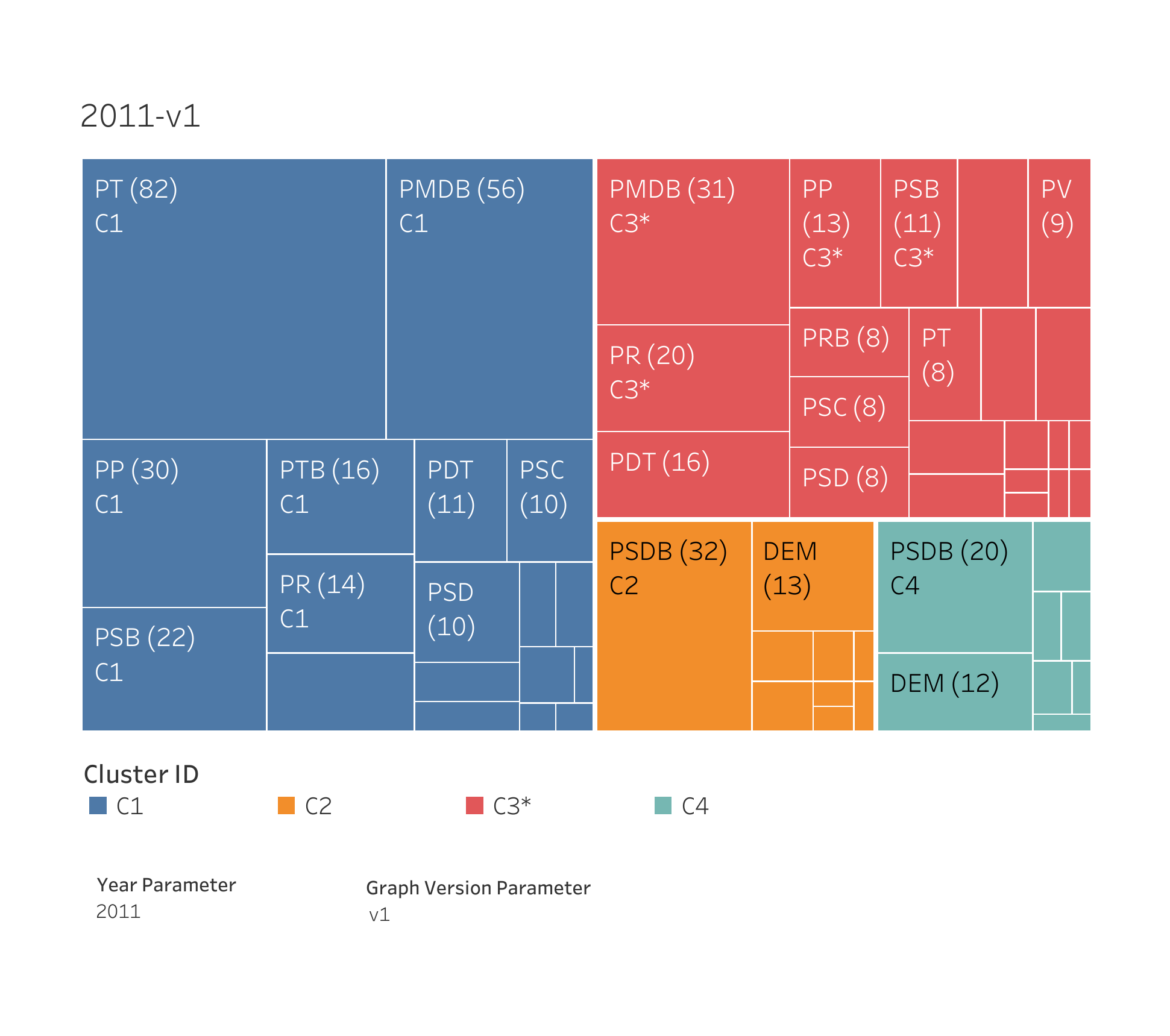}}
  %\vspace*{-1.4cm}
  \caption{SRCC clustering results for the year of 2011, when solving version v1 of the voting network, by fixing the number of clusters in the solution to $k = 4$. Each cluster is marked with a different color and a cluster label (begins with letter C, below the party name). For each cluster, the treemap displays the sum of deputies (in parenthesis), grouped by their respective party. Cluster labels marked with an asterisk (*) consist of mediation groups.}
  \label{fig:treemap-2011-v1}
  %\vspace*{-0.5cm}
\end{figure}

% Modelo para exibicao dos TreeMaps do Tableau
\begin{figure}[htbp]
  \centering
  \hspace*{-1.0cm}
  % Crop do PDF do Tableau
  % trim={<left> <lower> <right> <upper>}
  \adjustbox{trim={.05\width} {.2\height} {0.05\width} {.1\height},clip}%
  {\includegraphics[scale=0.75]{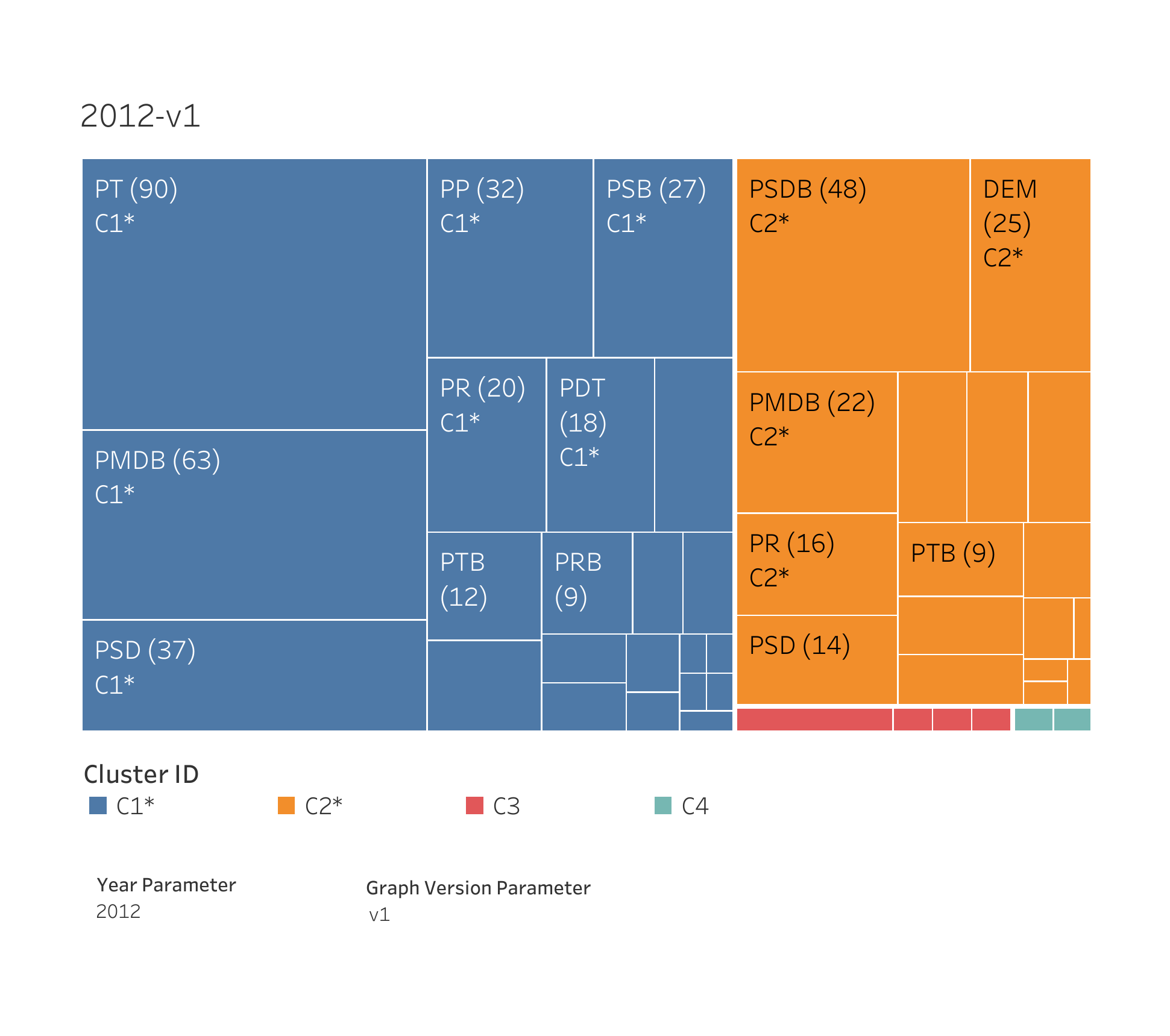}}
  %\vspace*{-1.4cm}
  \caption{SRCC clustering results for the year of 2012, when using version v1 of the voting network, by fixing the number of clusters in the solution to $k = 4$. Each cluster is marked with a different color and a cluster label (begins with letter C, below the party name). For each cluster, the treemap displays the sum of deputies (in parenthesis), grouped by their respective party. Cluster labels marked with an asterisk (*) consist of mediation groups.}
  \label{fig:treemap-2012-v1}
  %\vspace*{-0.5cm}
\end{figure}

We observe that, in the first year of president Dilma Rousseff’s government (2011), the government coalition is divided, roughly speaking, 
in two or three big groups, depending on the network version on which the analysis is based. According to version v1 (Figure~\ref{fig:treemap-2011-v1}), the largest cluster (C1) has 64\% of the allied deputies. Also, the great majority of the president’s party (PT), 82 deputies, are to be found in this cluster.

From 2012 onwards, a clear basis consolidation can be observed, with 77\% of the allied deputies in the same group (cluster C1 in Figure~\ref{fig:treemap-2012-v1}). 
%(C1 in both network versions, see Figure~\ref{fig:treemap-2012-v1} and ~\ref{fig:treemap-2012-v2}). 
%It represents the majority of the Chamber of Deputies (how much) and 
This cluster also holds more than 80 deputies of president’s party (PT). 

\begin{figure}[htbp]
  \centering
  \hspace*{-1.0cm}
  % Crop do PDF do Tableau
  % trim={<left> <lower> <right> <upper>}
  \adjustbox{trim={.05\width} {.2\height} {0.05\width} {.1\height},clip}%
  {\includegraphics[scale=0.75]{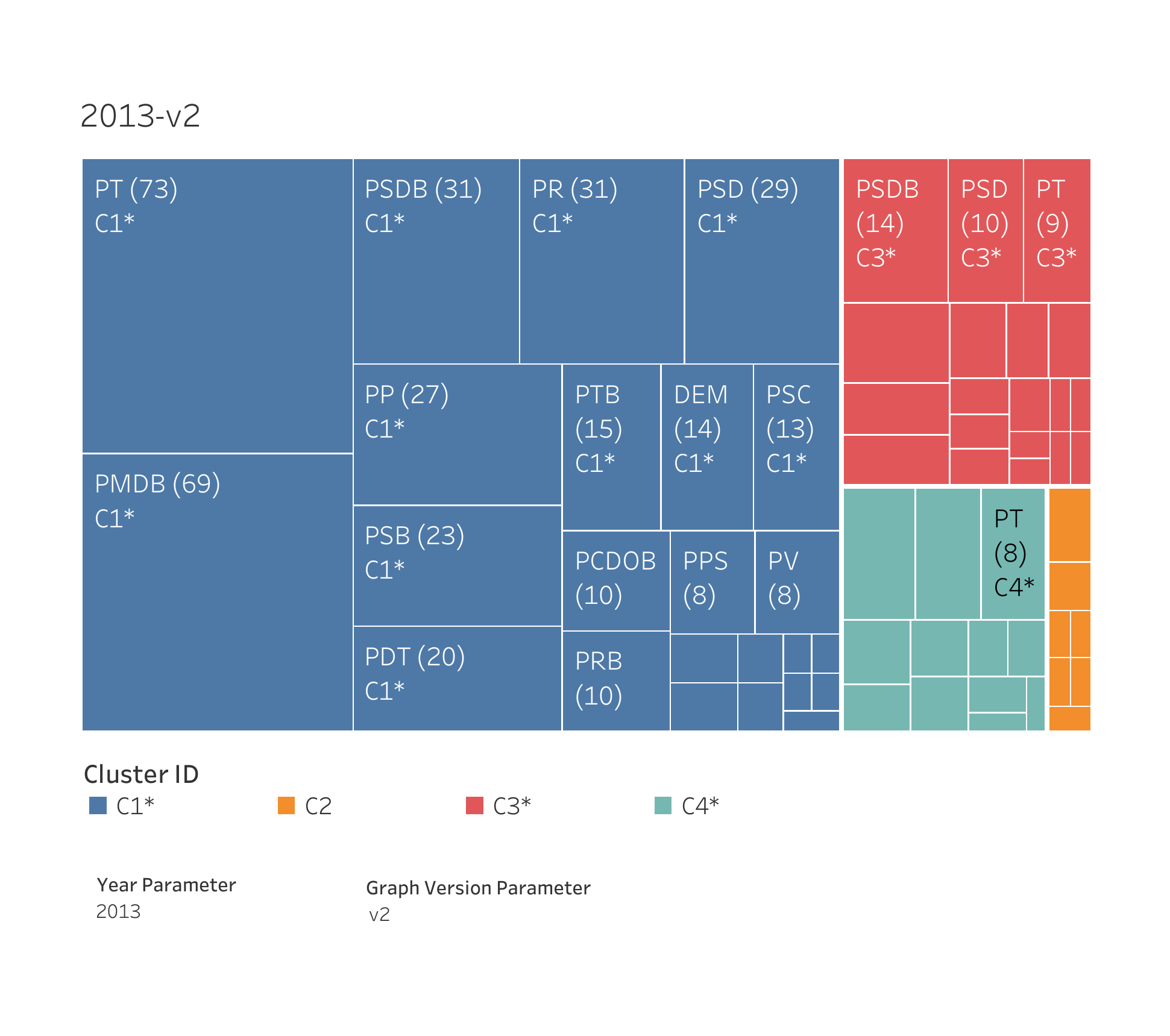}}
  %\vspace*{-1.4cm}
  \caption{SRCC clustering results for the year of 2013, when using version v2 of the voting network, by fixing the number of clusters in the solution to $k = 4$. Each cluster is marked with a different color and a cluster label (begins with letter C, below the party name). For each cluster, the treemap displays the sum of deputies (in parenthesis), grouped by their respective party. Cluster labels marked with an asterisk (*) consist of mediation groups.}
  \label{fig:treemap-2013-v2}
  %\vspace*{-0.5cm}
\end{figure}

\begin{figure}[htbp]
  \centering
  \hspace*{-1.0cm}
  % Crop do PDF do Tableau
  % trim={<left> <lower> <right> <upper>}
  \adjustbox{trim={.05\width} {.2\height} {0.05\width} {.1\height},clip}%
  {\includegraphics[scale=0.75]{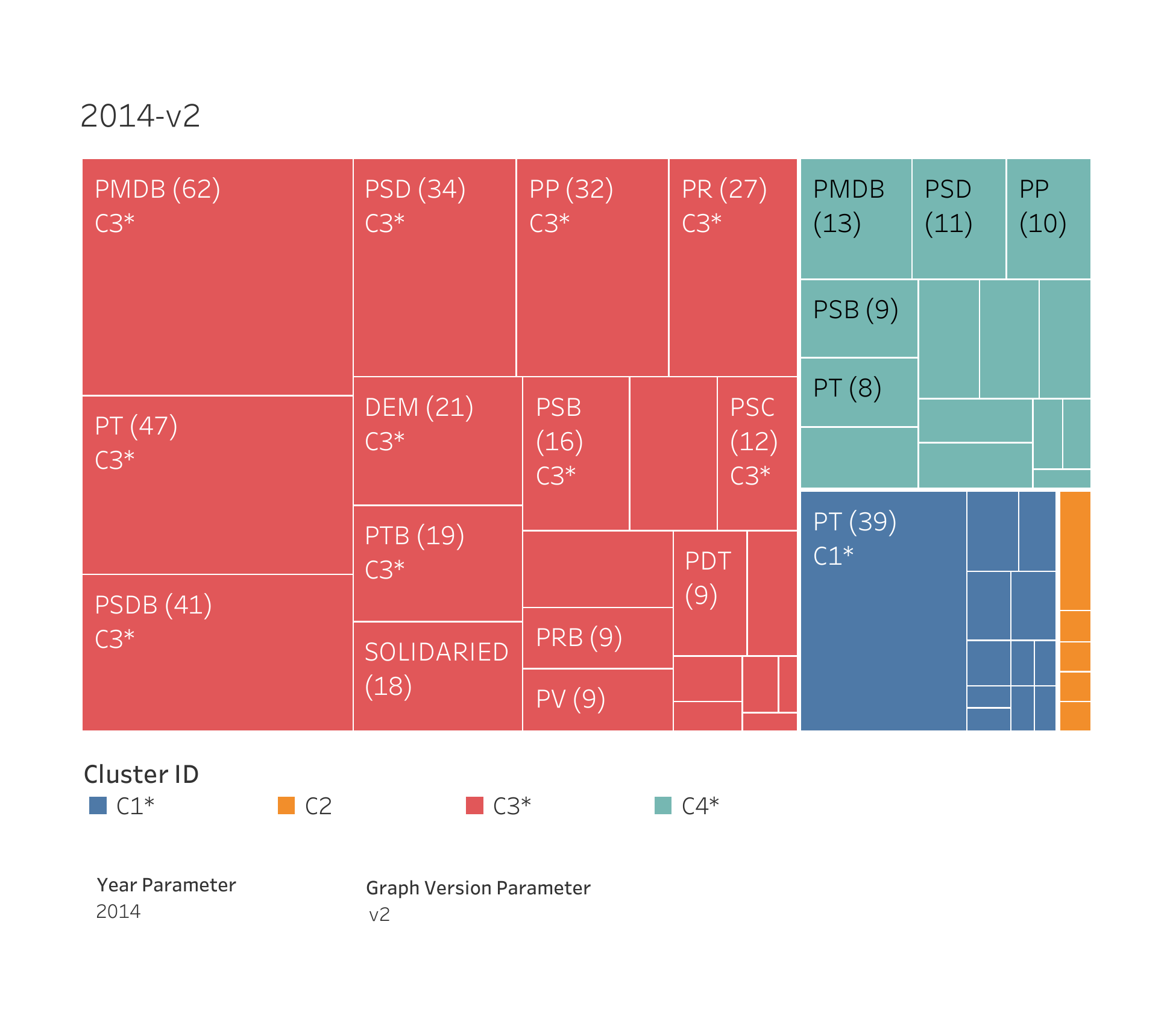}}
  %\vspace*{-1.4cm}
  \caption{SRCC clustering results for the year of 2014, when using version v2 of the voting network, by fixing the number of clusters in the solution to $k = 4$. Each cluster is marked with a different color and a cluster label (begins with letter C, below the party name). For each cluster, the treemap displays the sum of deputies (in parenthesis), grouped by their respective party. Cluster labels marked with an asterisk (*) consist of mediation groups.}
  \label{fig:treemap-2014-v2}
  %\vspace*{-0.5cm}
\end{figure}

In 2013 (Figure~\ref{fig:treemap-2013-v2}), the percentage of allied deputies inside the largest cluster (C1) rises to 82\% of the coalition (74 PT deputies). 
However, in 2014 (the last year of president Dilma Rousseff’s first term), a change of course comes about. 
This measure falls to 66\% (Figure~\ref{fig:treemap-2014-v2}) and, even worse, only about half of PT’s deputies are inside the main coalition group (C3). 

% Modelo para exibicao dos TreeMaps do Tableau

% Modelo para exibicao dos TreeMaps do Tableau

\begin{figure}[htbp]
  \centering
  \hspace*{-1.0cm}
  % Crop do PDF do Tableau
  % trim={<left> <lower> <right> <upper>}
  \adjustbox{trim={.05\width} {.2\height} {0.05\width} {.1\height},clip}%
  {\includegraphics[scale=0.75]{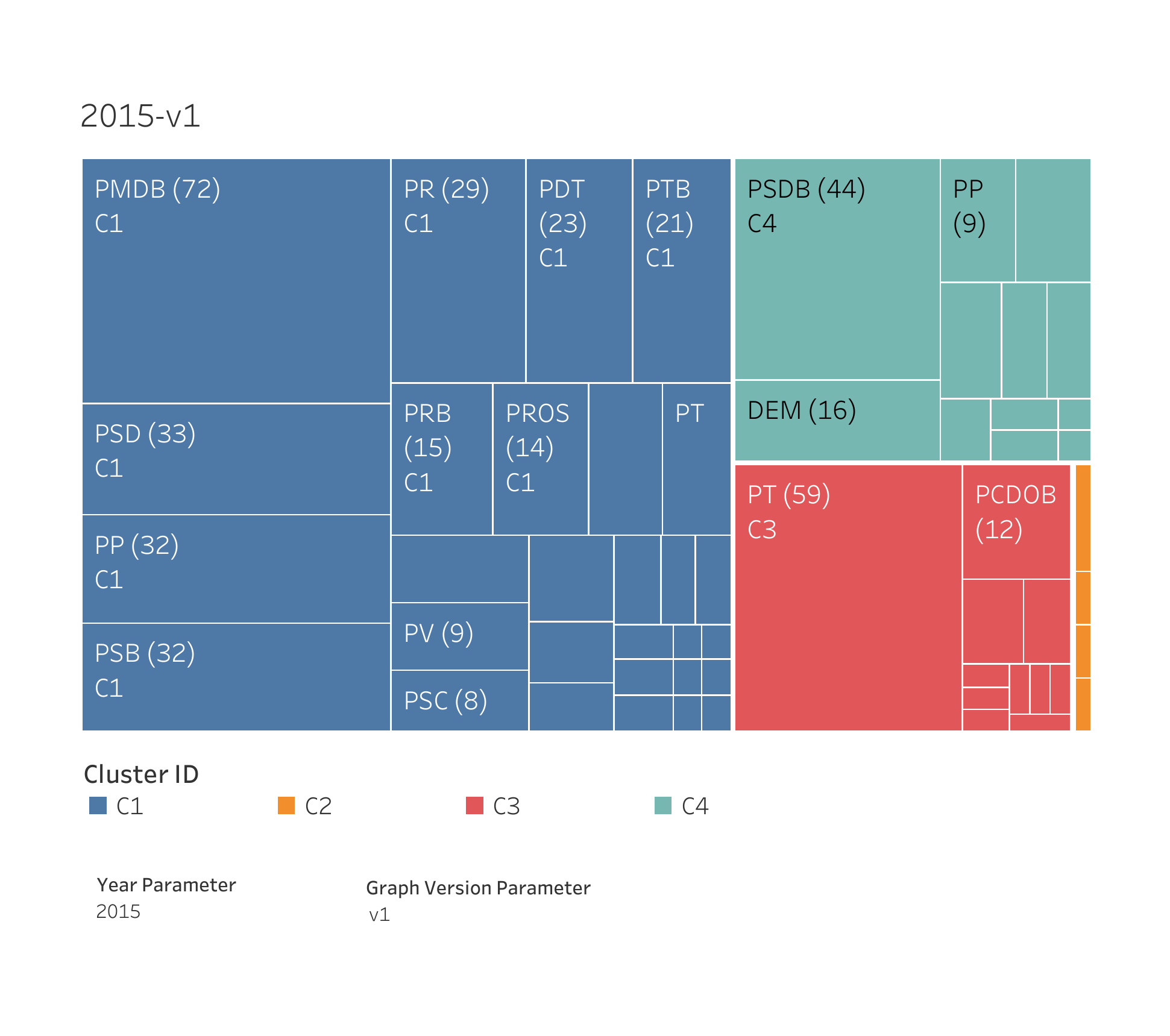}}
  %\vspace*{-1.4cm}
  \caption{SRCC clustering results for the year of 2015, when using version v1 of the voting network, by fixing the number of clusters in the solution to $k = 4$. Each cluster is marked with a different color and a cluster label (begins with letter C, below the party name). For each cluster, the treemap displays the sum of deputies (in parenthesis), grouped by their respective party. Cluster labels marked with an asterisk (*) consist of mediation groups.}
  \label{fig:treemap-2015-v1}
  %\vspace*{-0.5cm}
\end{figure}

A close look at president Dilma Rousseff’s second presidential term is surprising. 
In 2015, the biggest group of what should be the government’s new coalition (cluster C1 in Figure~\ref{fig:treemap-2015-v1}) is formed by 70\% of the total number of deputies of the coalition as a whole. Notwithstanding, this group houses at most 10 deputies of the president’s party (PT). Consider as well that the greatest part of PT deputies is in fact isolated in a smaller cluster, together with a few deputies from less influential parties. Note that both network versions show almost identical results\footnote{Please visit \url{https://public.tableau.com/profile/mario.levorato} for a full list of charts and tables.}. 

\begin{figure}[htbp]
  \centering
  \hspace*{-1.0cm}
  % Crop do PDF do Tableau
  % trim={<left> <lower> <right> <upper>}
  \adjustbox{trim={.05\width} {.2\height} {0.05\width} {.1\height},clip}%
  {\includegraphics[scale=0.75]{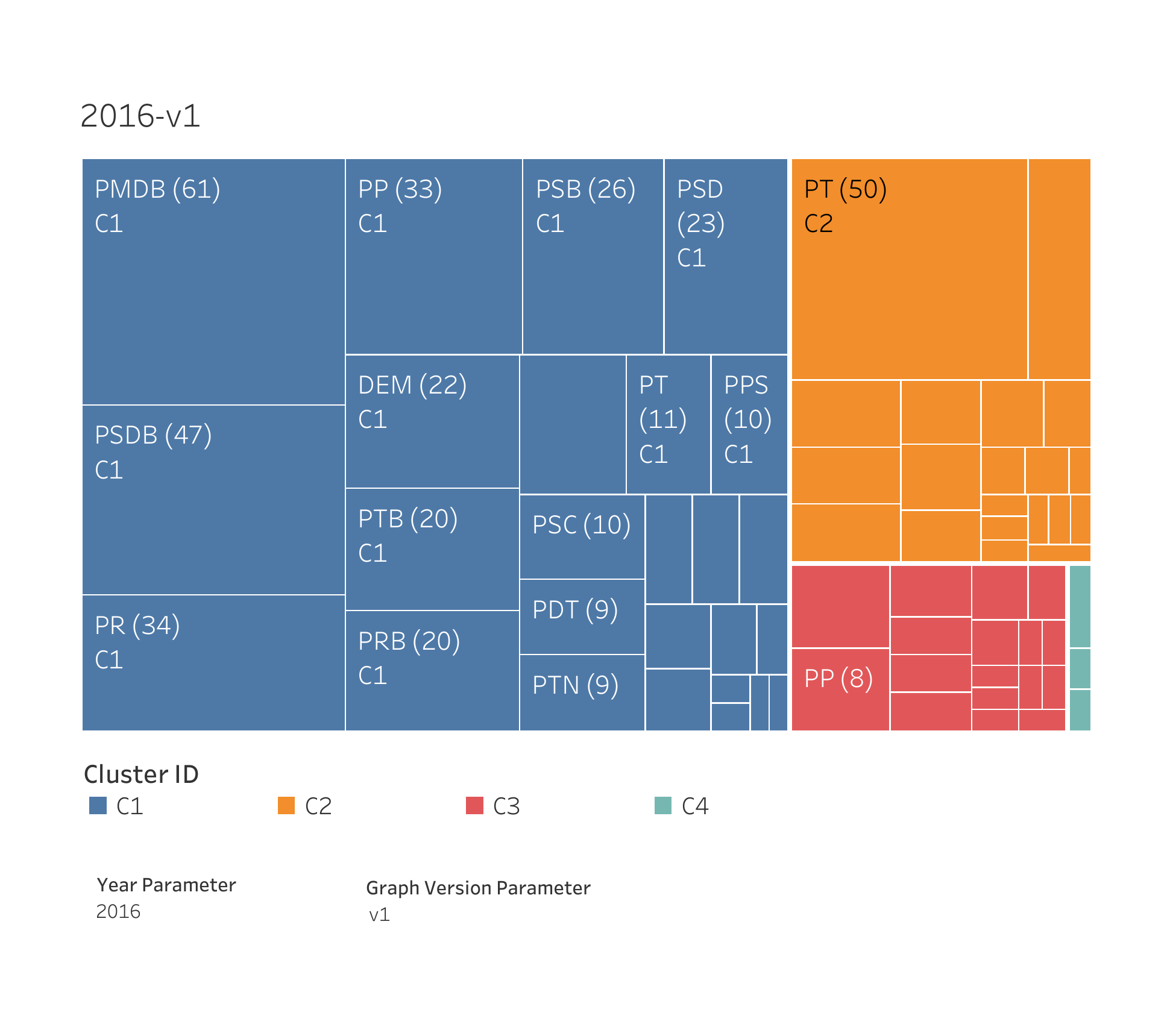}}
  %\vspace*{-1.4cm}
  \caption{SRCC clustering results for the year of 2016, when using version v1 of the voting network, by fixing the number of clusters in the solution to $k = 4$. Each cluster is marked with a different color and a cluster label (begins with letter C, below the party name). For each cluster, the treemap displays the sum of deputies (in parenthesis), grouped by their respective party. Cluster labels marked with an asterisk (*) consist of mediation groups.}
  \label{fig:treemap-2016-v1}
  %\vspace*{-0.5cm}
\end{figure}

A similar picture takes place in 2016 (Figure~\ref{fig:treemap-2016-v1}), when about two thirds of the supposedly allied deputies belong to the same group, 
which contains only 11 PT deputies. Similarly, 50 PT deputies can be found in another cluster.

Briefly speaking, results point out that in the years of 2015 and 2016, even though there are still large groups in which most deputies are from the so-called government coalition, such groups are no longer in accordance with the president’s party, which is perfectly understandable because of the political crisis and the loss of parliamentary support, news widely broadcast~\citep{dyer2015, boadle2016, watts2016a}. 

% Modelo para exibicao dos TreeMaps do Tableau

% Modelo para exibicao dos TreeMaps do Tableau

\subsection{Strength of party leadership}

This study was carried out as follows: for each year from 2011 to 2016 and for each party, we scanned data about the deputies and the clusters from which they make part. This information was then cross-referenced with the cluster where the leader of the respective party is found. This way it is possible to have a clear view of how strong the leadership of each party is: if a specific deputy belongs to different cluster than its party leader, on average, this deputy did not vote the way his party expected. The full results with the information about the deputies classified in the same cluster as their respective party leader (percentage) are available in Table~\ref{tab:PartyLeadership-2011-2014} for 2011-2014 and in Table~\ref{tab:PartyLeadership-2015-2016} for 2015-2016.

\begin{table}[htbp]
  \centering
  %\hspace*{-1.0cm}
  % Crop do PDF do Tableau
  % trim={<left> <lower> <right> <upper>}
  \adjustbox{trim={0.05\width} {0.05\height} {0.25\width} {.05\height},clip}%
  {\includegraphics[scale=1.05]{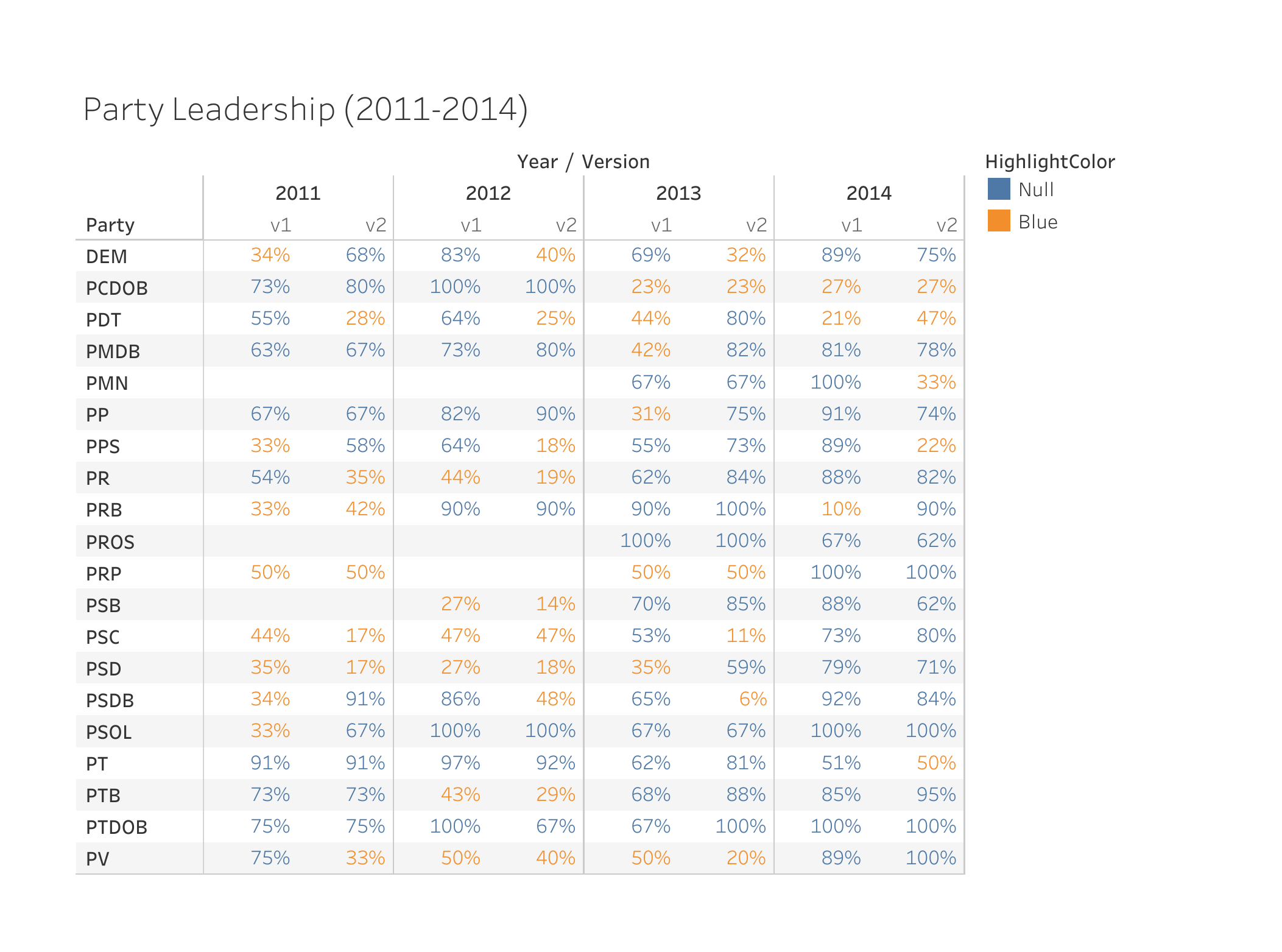}}
  %\vspace*{-1.4cm}
  \caption{For each party (column \textit{Party}), displays the percentage of its deputies who vote after their party leader (i.e. deputies classified in the same group of their party leader), for each year between 2011 and 2014 (columns \textit{2011} to \textit{2014}) and for each network version (columns \textit{v1} and \textit{v2}). On certain periods, the numbers associated with a party may not have been shown. Either because the party still did not exist at that time or did not have any representation in parliament at all.}
  \label{tab:PartyLeadership-2011-2014}
  %\vspace*{-0.5cm}
\end{table}

\begin{table}[htbp]
  \centering
  %\hspace*{-1.0cm}
  % Crop do PDF do Tableau
  % trim={<left> <lower> <right> <upper>}
  \adjustbox{trim={0.05\width} {0.05\height} {0.35\width} {.05\height},clip}%
  {\includegraphics[scale=1.05]{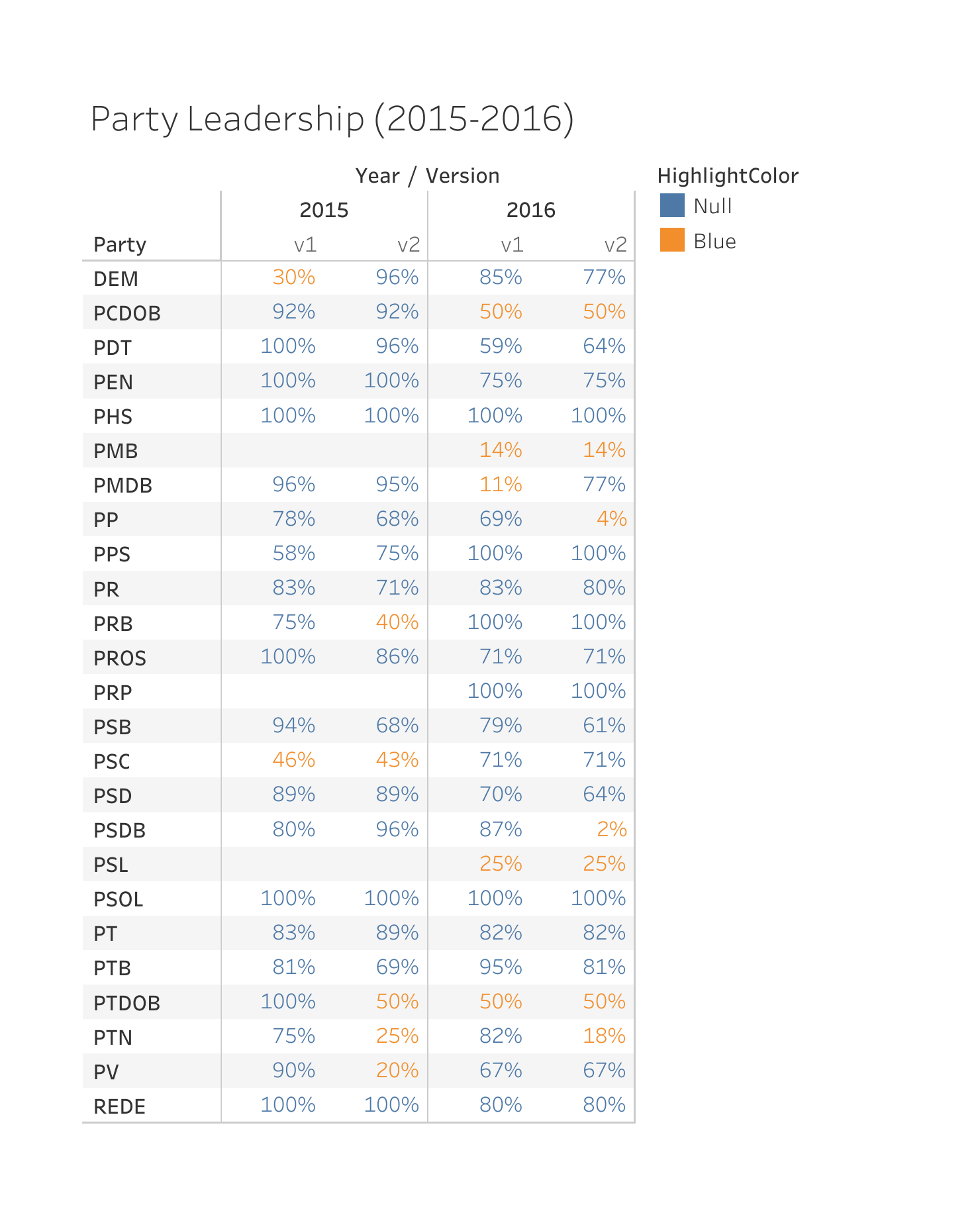}}
  %\vspace*{-1.4cm}
  \caption{For each party (column \textit{Party}), displays the percentage of its deputies who vote after their party leader (i.e. deputies classified in the same group of their party leader), for the years of 2015 and 2016 (columns \textit{2015} and \textit{2016}; until June 2016) and for each network version (columns \textit{v1} and \textit{v2}). On certain periods, the numbers associated with a party may not have been shown. Either because the party still did not exist at that time or did not have any representation in parliament at all.}
  \label{tab:PartyLeadership-2015-2016}
  %\vspace*{-0.5cm}
\end{table}

For each year, the following parties have been identified as having low percentage ($\rho < 50\%$) of deputies who vote after their party leaders, independently of the analyzed network version:

\begin{itemize}
    \item 2011: PRB, PRP, PSC, PSD.
    \item 2012: PR, PSB, PSC, PSD, PTB, PV.
    \item 2013: PCDOB, PRP, PV.
    \item 2014: PCDOB, PDT.
    \item 2015: PSC.
    \item 2016: PCDOB, PMB, PSL, PTDOB. 
\end{itemize}

Deep consideration into this list will reveal that, as we spot a considerably great number of deputies arranged in clusters where their party leaders are not present, there is strong evidence that, on average, voting recommendations  from party leaders have not been followed by many deputies.

\subsection{Detection of mediation groups in the chamber of deputies by the algorithm}

When the ILS-CC algorithm analyzes the voting networks of the Chamber of Deputies, two versions for each year, it proves to be quite successful in identifying positive mediation groups, that means, clusters whose most internal and external relationships are positive. In the conducted study, a group was classified as showing mediation properties whenever its positive relationship percentage was above 90\% and also its internal positive link ratio exceeded that same level. 

Because of the large number of political parties in the Brazilian CD (over 23 parties), we chose to rely our analysis on individual deputies instead of parties associated with mediation. 
We have cross-referenced the list of deputies inside each mediation cluster with the list of party leaders of the CD\footnote{The list of CD party leaders is available at \url{http://www2.camara.leg.br/deputados/liderancas-e-bancadas}.} plus the list of deputies that preside permanent committees of the CD\footnote{The list of permanent committees of the CD is available at \url{http://www2.camara.leg.br/atividade-legislativa/comissoes/comissoes-permanentes/}.}, for a total of 60 deputies who act as potential mediators.  Table~\ref{tab:mediation-clusters-and-qty-of-leaders} lists the mediation clusters, for each year and voting network version, as well as how many of its deputies belong either to party leadership or to permanent committees. Remark that in the years of 2015 and 2016 no mediation groups were detected, probably due to the unstable political environment. 
The obtained results suggest that the mediation groups detected by the SRCC algorithm are in accordance with the corresponding political scenario.  

% Acrescentar que em 2015 e 2016 não foram detectados grupos de mediação, devido ao ambiente político instável. Além disso, explicar que não é possível identificar partidos associados a mediação, devido à enorme quantidade de partidos. É mais fácil apontar deputados em separado, dos mais variados partidos.

Additionally, in every figure that contains a treemap (see Figure~\ref{fig:treemap-2014-v2}) showing what is inside each cluster (parties and number of deputies), the information about which groups have mediation properties is also present in the cluster labels marked with an asterisk (*). 

\begin{table}[htpb]
  \centering
  %\hspace*{-1.0cm}
  % Crop do PDF do Tableau
  % trim={<left> <lower> <right> <upper>}
  \adjustbox{trim={0.07\width} {0.05\height} {0.05\width} {0.23\height},clip}%
  {\includegraphics[scale=0.9]{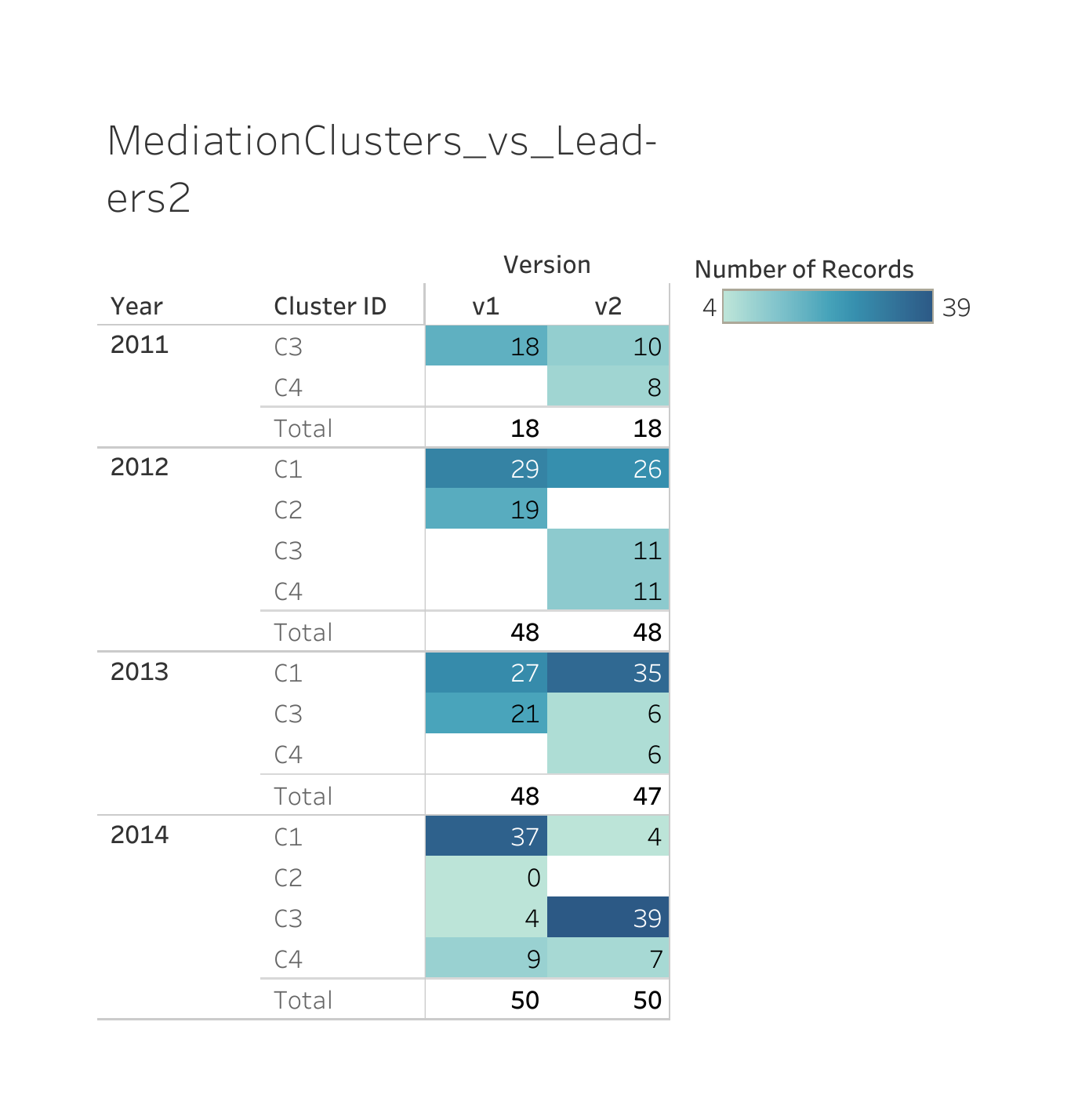}}
  %\vspace*{-1.4cm}
  \caption{For each year, network version (v1, v2) and mediation cluster detected by the algorithm, displays the quantity of deputies which belong either to party leadership or to permanent committees of the parliament. A blank cell means that mediation properties do not apply to the corresponding cluster for a specific year / network version.}
  \label{tab:mediation-clusters-and-qty-of-leaders}
  %\vspace*{-0.5cm}
\end{table}

%Did the president's political party (PT) split when the Brazilian political crisis began?
\subsection{The split of the government ruling party when the Brazilian political crisis began}

The clustering results for 2014 strongly suggest that president Dilma Rousseff’s party (PT) split, with 47 deputies in the first cluster, 39 in the second and 8 in the third. As seen on Figure~\ref{fig:treemap-2014-v2}, the treemap shows the fragmentation of PT in the last year of the president’s first term.

%After president Dilma Rousseff was reelected, did the government coalition loose support?
\subsection{Government coalition's loss of support after president Dilma Rousseff's reelection}

According to the results obtained by the ILS-CC algorithm, in 2015, after president Dilma Rousseff’s reelection, three clusters cover 99\% of the deputies (Figure~\ref{fig:treemap-2015-v1}). The parties inside each cluster reveal the main political groups at that time:
\begin{itemize}
    \item the largest group includes mainly center parties, such as the majority of PMDB, PSD, PP and PR;
    \item the second biggest group is formed by opposition parties like PSDB and DEM;
    \item the last one represents the government core parties, such as PT (59 deputies) and PCDOB (12 members). 
\end{itemize}

A comparison between 2015 and previous years (see Figures~\ref{fig:treemap-2011-v1}, \ref{fig:treemap-2012-v1} and \ref{fig:treemap-2015-v1}) reveals that the government coalition has gone through a substantial loss of support, mostly from center parties. These results anticipate a movement which became clear only the following year, when PMDB and other center-parties voted to leave the governing coalition~\citep{boadle2016, watts2016a, barchfield2016}. 

%When the government coalition lost power, did the center parties move towards opposition?
\subsection{Center parties moved towards opposition when the government coalition lost power}

Looking at the data for the years of 2015 (Figure~\ref{fig:treemap-2015-v1}) and 2016 (Figure~\ref{fig:treemap-2016-v1}), one can observe that the majority of center party and opposition deputies started sharing the same group. There was a strong approximation between PMDB (center), PSDB and DEM (opposition), which have previously been in separate clusters. According to the charts, one can notice that center parties have moved towards opposition. 

In 2015, there was a large movement of parties from the government coalition, which went to a “super-centered” group. These parties include: PROS (12), PRB (12), PDT (22), PR (25), PP (28), PSD (33) and PMDB (71).

In 2016, the following coalition parties have effectively migrated to what can be interpreted as a huge opposition cluster: PDT(17), PRB (20), PSD (21), PP (30), PR (33) and PMDB (56). 

News broadcast confirm this movement: first, the approximation between Brazil's biggest party (PMDB) and PSDB was reported~\citep{gonalves2016, sambo2016}. Shortly after, PMDB voted to leave the governing alliance~\citep{boadle2016, watts2016a}, followed by three other parties (PDT, PRB and PP)~\citep{barchfield2016}. 

%At any time, do numbers show polarization between political groups?
\subsection{Polarization between political groups}

In 2012 (on both network versions), the chamber of deputies is polarized in two large groups (see Figure~\ref{fig:treemap-2012-v1}). The first one with 238 members, led by the majority of PT and PMDB deputies (government base). The other cluster is mainly characterized by opposition parties, such as PSDB and DEM, but it also includes dissidents from center parties like PMDB and PSD. 

\subsection{Relative imbalance of the analyzed signed social networks}

Several authors have mentioned that real-world signed social networks are more balanced than expected~\citep{Kunegis09, Leskovec10, kunegis2010spectral, Facchetti11}. As seen on Table~\ref{tab:sri}, the signed social networks generated from the Brazilian CD voting data are in fact highly balanced, which supports existing research about that topic.

\begin{table}[htbp]
\begin{center}
\begin{tabular}{|l|c|c|c|c|c|c|cc|}
\hline
\textbf{Year} & \multicolumn{ 2}{c|}{\textbf{2010}} & \multicolumn{ 2}{c|}{\textbf{2011}} & \multicolumn{ 2}{c|}{\textbf{2012}} & \multicolumn{ 2}{c|}{\textbf{2013}} \\ \hline
\textbf{Version} & v1 & v2 & v1 & v2 & v1 & v2 & \multicolumn{1}{c|}{v1} & \multicolumn{1}{c|}{v2} \\ \hline
\textbf{$\% SRI(P)$} & 0.25\% & 0.26\% & 0.38\% & 0.43\% & 0.34\% & 0.33\% & \multicolumn{1}{c|}{0.31\%} & \multicolumn{1}{c|}{0.35\%} \\ \hline \hline
\textbf{Year} & \multicolumn{ 2}{c|}{\textbf{2014}} & \multicolumn{ 2}{c|}{\textbf{2015}} & \multicolumn{ 2}{c|}{\textbf{2016}} &  &  \\ \cline{1-7}
\textbf{Version} & v1 & v2 & v1 & v2 & v1 & v2 &  &  \\ \cline{1-7}
\textbf{$\% SRI(P)$} & 0.07\% & 0.07\% & 0.39\% & 0.40\% & 1.92\% & 2.32\% &  &  \\ \hline
\end{tabular}
\end{center}
\caption{Symmetric Relaxed Imbalance ( $\%SRI(P)$ ) measure obtained with the solution of the SRCC problem over the CD signed graphs, according to year and network version.}
\label{tab:sri}
\end{table}

%% file: conclusion.tex
In this article, we have investigated some of the aspects
inherent to signed voting networks and political relationships, by using data from the Brazilian Chamber of Deputies (CD). We have first extracted a collection of networks based on voting patterns of the CD members. We have also applied a clustering algorithm specifically designed for signed networks, called $ILS-CC$, which aims to reduce structural balance.

The analysis of the identified clusters has shown that mediation groups do exist in the Chamber of Deputies. Detected in different years, these groups include several deputies who might act as potential mediators, such as party leaders and presidents of permanent committees of the CD. The applied algorithm has also allowed us to gather evidence that certain parties are indeed unfaithful to their coalition. Besides, the obtained data perfectly confirms the news broadcast about the Brazilian political situation, such as the loss of support that government coalition experienced. 

Equally, the algorithm has proved to be a useful tool to spot parties under weak leadership and the existence of polarization between two large political groups. Our analysis also confirmed that the signed social networks we generated from the Brazilian CD voting data are indeed extremely balanced, hence supporting previous related works.